\documentclass[journal]{IEEEtran}
\usepackage{amsmath,amsfonts}
\usepackage{amssymb}
\usepackage{algorithmic}
\usepackage{algorithm}
\usepackage{array}
\usepackage[caption=false,font=footnotesize,labelfont=rm,textfont=rm]{subfig}
\usepackage{textcomp}
\usepackage{stfloats}
\usepackage{url}
\usepackage{verbatim}
\usepackage{graphicx}
\usepackage{cite}
\usepackage{makecell}
\usepackage{booktabs}
\usepackage{color}
\usepackage{hyperref}

\newtheorem{theorem}{Theorem} 
\begin{document}

\title{Cost-Effective XL-MIMO Communication with Cylinder Directly-Connected Antenna Array}%Directly-Connected Antenna Array: Cylinder Architecture with Conformal Semi-Circular array for Enhanced Wireless Communication Performance}

\author{
	Xuancheng Zhu, ~\IEEEmembership{Student Member,~IEEE,} Zhiwen Zhou, Zhenjun Dong,  ~\IEEEmembership{Student Member,~IEEE,} and 
	Yong Zeng, ~\IEEEmembership{Fellow,~IEEE}
	\thanks{
    Part of this work will be presented at 2025 IEEE Globecom, the 6th workshop on emerging topics in 6G communications, 2025. \cite{ref:omnicell}
    
    X. Zhu, Z. Zhou, and Y. Zeng are with the National Mobile Communications Research Laboratory, Southeast University, Nanjing 210096, China. Y. Zeng is also with the Purple Mountain Laboratories, Nanjing 211111, China (e-mail: \{213223563, zhiwen\_zhou, yong\_zeng\}@seu.edu.cn). Z. Dong is with the Purple Mountain Laboratories, Nanjing 211111, China (e-mail: dongzhenjun@pmlabs.com.cn). (Corresponding author: Yong Zeng.)
    } 
    }

\maketitle

\begin{abstract}
Extremely large-scale multi-input multi-output (XL-MIMO) is a promising technology for the sixth generation (6G) wireless networks, thanks to its superior spatial resolution and beamforming gains.  
In order to realize XL-MIMO cost-effectively, an innovative \emph{ray antenna array} (RAA) architecture with directly-connected \emph{uniform linear array} (ULA) was recently proposed, which achieves flexible beamforming without relying on traditional analog phase shifters or digital beamforming.  
However, RAA suffers from the signal blockage issue since its ray-configured ULAs are placed in the same plane. 
To address this issue, this paper proposes a novel antenna array architecture termed cylinder \emph{directly-connected antenna array} (DCAA), 
% with enhanced wireless communication performance for the high-frequency band such as millimeter wave (mmWave) and Terahertz (THz) in the sixth-generation (6G) communication. 
which is achieved via multiple \emph{simple uniform circular array} (sUCA) with carefully designed orientations in a layered three-dimensional structure. 
The so-called sUCA partitions the \emph{uniform circular array} (UCA) into two sub-arrays where each sub-array has all antenna elements directly connected to achieve a desired beam direction corresponding to the sub-array's physical orientation, thus achieving full spatial coverage. 
Compared with the conventional ULA architecture with hybrid analog/digital beamforming (HBF), the proposed cylinder DCAA can achieve uniform spatial resolution, enhanced communication rate and lower hardware costs. 
Simulation results are provided to validate the promised gains of cylinder DCAA, demonstrating its great potential for high-frequency systems such as millimeter
wave (mmWave) and Terahertz (THz) systems.

\end{abstract}


\section{Introduction}
% {\color{blue} Over the past few decades, wireless communication has undergone remarkable advancements, transforming the way information is transmitted and received. 
% One of the most significant breakthroughs is the development of multi-input multi-output (MIMO) technology, which uses multiple antennas to achieve the spatial diversity, multiplexing and beamforming gains, thus dramatically improving communication capacity, reliability, and spectral efficiency \cite{wireless_comm}.} 
\IEEEPARstart{O}{ver} the past few decades, multi-input multi-output (MIMO) technology has served as a cornerstone in the evolution of wireless communications \cite{wireless_comm}.
% {\color{blue}
% The evolving MIMO technology heads for increasing antenna elements in the large scale array and innovative structure to enhance communication performance and improve cost-efficiency to achieve future green communication. }
From the fourth-generation (4G) to the fifth-generation (5G) mobile communication networks, multi-antenna technology has evolved from MIMO (typically with 8 antenna elements) to massive MIMO (typically with 64 antenna elements), to substantially enhance spectral efficiency, reliability and connection density. 
For the sixth-generation (6G) mobile communication networks, more ambitious goals have been envisioned, such as centimeter-level positioning accuracy, ultra-high connection density, and ultra-low latency on the order of 0.1–1 ms \cite{ref:6G,ref:6G_sensing}. 
% {\color{blue} Since it is generally believed that 6G will utilize higher frequency bands to achieve these ambitious visions, such as millimeter wave (mmWave) and TeraHertz (THz) communication, MIMO technology with extremely large number of antenna elements is realizable within limited physical size for 6G network. 
% Therefore, extremely-large scale MIMO (XL-MIMO) with hundreds or even thousands of antenna elements was recently proposed as an effective option to achieve requirements of 6G network. Attributing to the extremely large number of antenna elements, XL-MIMO achieves higher beamforming gain, more compact beam lobe and more precise sensing and localization.}  
To meet these goals, various new MIMO technologies, such as extremely-large scale MIMO (XL-MIMO) \cite{ref:nearfield}, sparse MIMO \cite{ref:sparse_MIMO} and cell-free MIMO \cite{ref:cell_free} are being actively explored. In particular, attributing to the extremely large number of antenna elements, XL-MIMO achieves higher beamforming gain, better spectral efficiency and more precise sensing and localization ability. 
With the growing trend of utilizing higher frequency bands for 6G networks, such as millimeter wave (mmWave) and Terahertz (THz), the practical implementation of XL-MIMO faces significant challenges, including the high hardware costs associated with pricey front-end radio frequency (RF) components, increased signal processing complexity and the difficulty of designing and fabricating advanced components, such as high-precision phase shifters. 
These factors collectively hinder the practical implementation of XL-MIMO. 
% However, with the trend of utilizing higher frequency bands, such as millimeter wave (mmWave) and Terahertz (THz), larger array scale and wider spectral resources for 6G communication, existing XL-MIMO technology deployed on extremely-large spectral space with pricey front-end radio frequency (RF) components, which are difficult to design and fabricate, is challenging to be practically implemented. 
% can not meet the vision of green communication for 6G network. 
Therefore, it is necessary to develop innovative and cost-effective antenna array structures.
Since RF chains and front-end components account for a substantial portion of hardware costs \cite{ref:shifter,ref:switch} and power consumption, 
% XL-MIMO systems inevitably suffer from significantly degraded cost-efficiency. 
an effecitve strategy to realize XL-MIMO is to reduce the number of RF chains or adopt more efficient front-end components \cite{ref:effect_comp}. 
For instance, hybrid analog/digital beamforming (HBF) \cite{ref:HBF}, lens antenna array \cite{ref:lens}, movable antenna array \cite{ref:movable_1} or fluid antenna systems \cite{ref:fluid_antenna} and reconfigurable tri-hybrid MIMO antenna array \cite{ref:reconfigurable} have been investigated. 
% which is realized by the front-end analog beamforming and the back-end digital baseband beamforming. 
%which provides a more flexible trade-off between hardware cost and performance by using much fewer RF chains than antenna elements. 
%However, if the HBF or reconfigurable tri-hybrid MIMO antenna array architecture needs to be applied to achieve $360^\circ$ coverage, the classic cell sectoring ULA serves a cell sector with $120^\circ$ angular coverage. 
However, for HBF and reconfigurable tri-hybrid MIMO, the required number of phase shifters increases with the number of antennas, which is huge for XL-MIMO systems. 
Besides, HBF requires complicated control systems for accurate and dynamic phase control, thus increasing hardware complexity and power consumption. 
For lens antenna, dielectric lenses are difficult to integrate with multiple antenna techniques due to their bulky size and high insertion loss \cite{ref:lens_cons}. 
% For lens antenna array, extra power consumption is required to generate high-gain beams compensating for the large free-space loss, atmospheric absorption and weather effects \cite{ref:lens_cons}. 
For XL-MIMO based on movable antenna or fluid antenna array, accurate channel state information (CSI) is challenging to be estimated \cite{ref:movable_cons}. 
Besides, extra bulky mechanical controls need to be introduced, which compromises the cost-efficiency, power consumption and response time. 

\begin{figure*}[htbp]
    \centering
    \includegraphics[width=\linewidth]{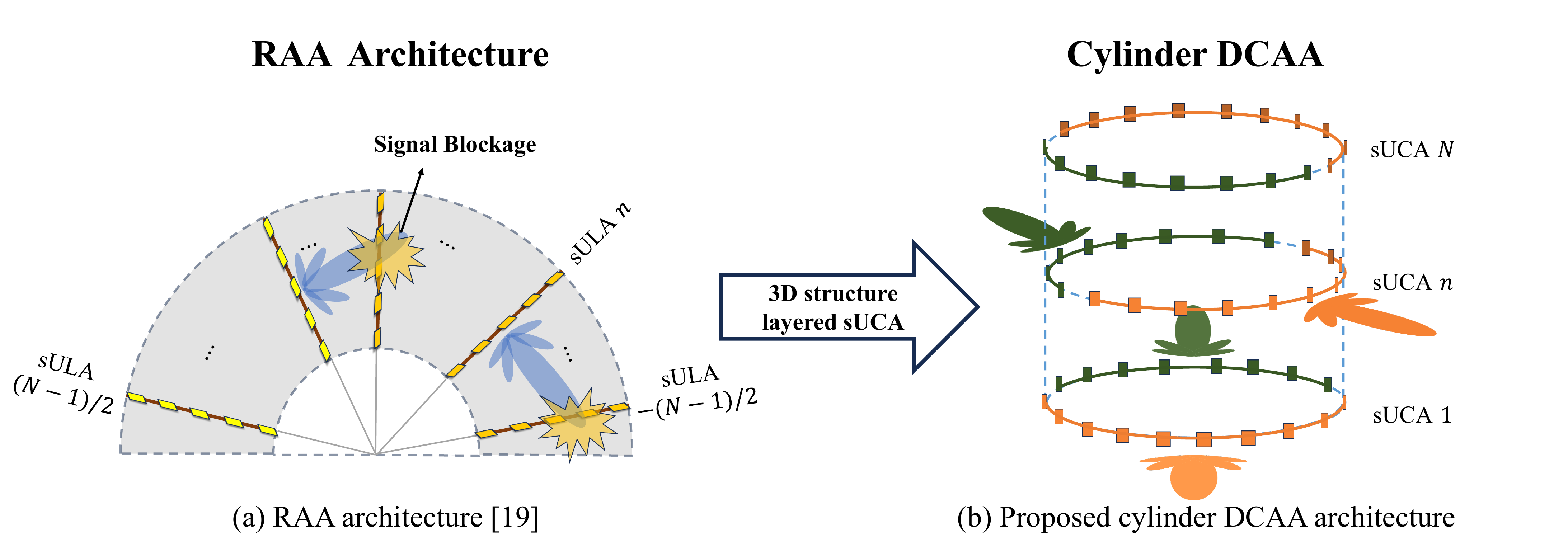}
    \caption{RAA architecture and the proposed cylinder DCAA architecture, which addresses the signal blockage issue by arranging multiple sUCAs in a layered 3D structure.}
    \label{fig:signal_blockage}
\end{figure*}

To address the above issues, an innovative multi-antenna architecture termed \emph{ray antenna array} (RAA) was recently proposed, which offers a promising solution to simultaneously reduce hardware cost and enhance system performance \cite{RAA,RAA_long}. As illustrated in Fig. \ref{fig:signal_blockage}(a), the RAA architecture consists of a large number of low-cost antenna elements arranged into a ray-like structure, where each ray corresponds to a \emph{simple uniform linear array} (sULA) with a carefully designed orientation. All antenna elements within each sULA are directly connected, so that each sULA is able to form a beam aligned with its physical orientation.   
%communication structure termed omnicell communication system \cite{ref:omnicell} is proposed, and it has the BS equipped with DCAA to address cost-efficiency problem and enhance communication performance. 
%Ray Antenna Array (RAA) \cite{RAA,RAA_long}, as a kind of DCAA proposed to be implemented in the omnicell communication system, uses multiple directly-connected ULAs termed \emph{simple uniform linear arrays} (sULAs) that connects all antenna elements directly to a specific RF chain without any beamforming technology. % so that no expensive phase shifters are needed in the architecture. 
%The RAA is inspired by the observation that without any analog or digital beamforming, a sULA can still form a beam in the direction matching its physical orientation. 
With the assistance of a ray selection network (RSN), RAA can achieve uniform spatial resolution, enhanced beamforming gain, high-precision sensing and reduced hardware costs \cite{RAA_long, ref:ISAC_with_RAA}, though at the expenditure of more bulky array size. 
However, when multiple sULAs are deployed on the same two-dimensional (2D) plane with different orientations, the RAA architecture may suffer from signal blockage issue among different sULAs. 
Specifically, the beam generated by a sULA propagates directly toward its forward direction, where adjacent sULAs are located, thereby physically obstructing the transmitted signals, as illustrated in Fig. \ref{fig:signal_blockage}(a).

% \begin{figure}[!t]
% 	\centering
% 	\includegraphics[width=\linewidth]{cylinder_system.pdf}
% 	\caption{The scheme of omnicel comunication system where BS is equipped with cylinder DCAA. }
% 	\label{fig:cylinder_system}
% %	\subfloat[BS equipped ULA-based cell sectoring with HBF]{%
% %		\includegraphics[height=3.2cm]{cell_sectoring.pdf}
% %		\label{fig:HBF}
% %	}
% %	\hspace{0.3cm}
% %	\subfloat[Proposed Cylinder architecture uplink scenario.]{%
% %		\includegraphics[height=3.2cm]{system.pdf}
% %		\label{fig:cylinder_system}
% %	}
% %	\caption{System architecture of HBF and proposed cylinder antenna array. The innovative architecture in (b) achieves less analog phase shifters than HBF, which decrease the hardware cost and improve the cost-efficiency.} 
% %	\label{fig:sys_compare}
% \end{figure}

% \begin{figure*}
% 	\centering
% 	\includegraphics[width=\linewidth]{sys_comparison.pdf}
% 	\caption{
% 		An overall comparison of cell sectoring system based on ULA with HBF and omnicell system based on RAA and proposed cylinder DCAA. Their basic components and beam steering strategies are different from each other. 
% 		%Systematic comparison of cell sectoring and omnicell communication system. 
% 	}
% 	\label{fig:sys_comparison}
% \end{figure*}

To address the aforementioned signal blockage issue in the RAA architecture, this paper proposes a novel antenna array design, referred to as cylinder \emph{directly-connected antenna array} (DCAA). 
As illustrated in Fig. \ref{fig:signal_blockage}(b), the proposed cylinder DCAA arranges multiple differently orientated \emph{simple uniform circular arrays} (sUCAs) into a three-dimensional (3D) layered structure. 
The so-called sUCA partitions the antenna elements equally into two complementary semi-circular-configured sub-arrays, and each sub-array has the antenna elements directly connected with the variable length antenna delay line technology. 
% As illustrated in Fig. \ref{fig:signal_blockage}, the proposed cylinder DCAA that arranges multiple sUCAs in a compactly layered three-dimensional (3D) structure is a potential solution for the signal blockage problem, due to the fact that a sUCA in the cylinder DCAA will not block the signal from other sUCAs. 
Through the selection matrix, sub-arrays of all sUCAs in the cylinder DCAA are selectively connected with few RF chains for further digital baseband processing. 
Thanks to the 3D layered structure, the cylinder DCAA can effectively avoid physical blockage among arrays, thereby resolving the signal blockage issue inherent in the RAA architecture. 
%Attributing to the 3D layered structure of cylinder DCAA, the arrays physical blockage is eliminated such that cylinder DCAA addresses the signal blockage issue of RAA structure. 
% Attributing to all antenna elements within each sub-array directly connected to a specific RF chain through the selection matrix instead of plenty of phase shifters, the cylinder DCAA will decrease the hardware costs and improve the quality of service (QoS) for the future 6G communication. 
% The overall system comparison of cell sectoring system based on ULA with HBF and omnicell system based on RAA and proposed cylinder DCAA is illustrated in Fig. \ref{fig:sys_comparison}. 
%maximum sum rate performance for dense connectivity as well. 
The main contributions of this paper are summarized as follows. 

\begin{itemize}
    \item 
    First, we present the overall system model of cylinder DCAA in Section \ref{sec:sys_model} and the geometric structure and the array response pattern of sUCA in Section \ref{sec:sUCA}, which is the basic component of the proposed cylinder DCAA. 
    Composed of two complementary sub-arrays, the sUCA incorporates the direct RF connections utilizing variable length antenna delay lines for the antenna elements within each sub-array to achieve desired beam directions without analog beamforming technology.
    % The geometric model of sUCA consisting of two sub-arrays accounts for the direct RF connection of the antenna elements in a sub-array, which is designed to achieve desired beam lobes by employing variable-length antenna delay lines. 
    We provide a comprehensive mathematical analysis of the response pattern of sUCA's sub-arrays with any given orientation. 
    The study demonstrates that the sUCA's sub-arrays achieve desired spatial resolution and low side-lobe levels, making them suitable for constructing more complex array structures.
    \item 
    Second, the detailed parameters design of the proposed cylinder DCAA is derived in Section \ref{sub_sec:define} and \ref{sub_sec:design} based on the characteristics of sUCA. 
    The cylinder DCAA is constructed by stacking multiple sUCAs vertically into a layer-configured structure, with each sUCA consisting of two complementary sub-arrays with carefully designed orientations to achieve uniform spatial resolution and avoid physical blockage among sub-arrays. 
    Since the antenna elements within each sub-array of sUCAs are directly connected without using any phase shifters, the proposed cylinder DCAA significantly reduces hardware costs and implementation difficulty compared to the classic HBF architectures. 
    The resulting cylinder DCAA provides enhanced beamforming gains, uniform spatial coverage, and improved cost-efficiency, making it a promising solution for flexible beamforming, especially for high frequency systems like mmWave or THz systems in 6G networks. 
    \item 
    Finally, detailed input-output signal models for the proposed cylinder DCAA in both uplink and downlink communication scenarios are derived in Section \ref{sec:signal_model}. 
    We evaluate the communication performance of the cylinder DCAA in terms of signal-to-interference-plus-noise ratio (SINR) and maximum sum rate, comparing it with a cell sectoring system based on ULA with HBF. 
    Simulation results in Section \ref{sub_sec:simu} and \ref{sub_sec:eff_and_conv} reveal that the proposed cylinder DCAA achieves superior communication rates, particularly in dense connectivity scenarios, thanks to the uniform spatial resolution. 
    Additionally, a cost analysis in Section \ref{sub_sec:cost} highlights that the cylinder DCAA is more cost-efficient than cell sectoring based on ULA with HBF, making it a practical and scalable solution for future 6G networks. 
\end{itemize}

{\it Notation}:
Italic, bold-faced lower- and upper-case letters denote scalars, vectors, and matrices, respectively.
The transpose and Hermitian transpose operation are given by $(\cdot)^T$ and $(\cdot)^H$, respectively.
$\mathbb{C}^{M\times N}$ and $\mathbb{R}^{M\times N}$ signify the spaces of $M\times N$ complex and real matrices. 
$\mathbf{diag}(\mathbf{a})$ denotes a diagonal matrix with its diagonal entries composed from the vector $\mathbf{a}$ in sequence. 
$\mathbb{Z}$ denotes the space of integers. 
$\textbf{1}_{M\times N}$ represents an $M\times N$ matrix with all entries equal to 1.
$j=\sqrt{-1}$ denotes the imaginary unit of complex numbers.
The distribution of a circularly symmetric complex Gaussian (CSCG) random variable with mean 0 and variance $\sigma^2$ is denoted by $\mathcal{CN}(0,\sigma^2)$, and $\mathcal{N}(0,\sigma^2)$ denotes the real-valued Gaussian distribution.
$\mathcal{U}(a,b)$ denotes the continuous uniform distribution at the interval $[a,b]$. 
$\vert\cdot\vert$ denotes the absolute value of a scalar. 
%${\rm {round}}(\cdot)$ represents rounding to the nearest integer.
%$\vert\cdot\vert_0$ denotes the $l_0$ norm.
$\vert\vert\cdot\vert\vert$ denotes the $l_2$ norm of a vector.
$\lceil\cdot\rceil$ and $\lfloor\cdot\rfloor$ denote the ceiling and floor operations, respectively. $\odot$ represents the Hadamard product. 
$\mathbf{sup}\{\cdot\}$ and $\mathbf{inf}\{\cdot\}$ denote the supremum and the infimum of a set, respectively. 
$\angle(\cdot)$ denotes the main phase of a complex scalar. 
$\mathbf{Re}(\cdot)$ and $\mathbf{Im}(\cdot)$ denote the real component and the imaginary component of a complex scalar, respectively. 
$(\cdot)^*$ denotes the conjugate of a complex scalar or vector. 
$\mathbb{E}[\cdot]$ denotes the mathematical exception of a random scalar. 
%$\otimes$ and $\odot$ denote the Kronecker and Hadamard product operations, respectively.
%$\otimes$ denotes the Kronecker product operation.
%$\text{vec}(\cdot)$,  $\text{diag}(\cdot)$, and $\text{tr}(\cdot)$ denote the vectorization operation, diagonalization operation, and the trace of a square matrix, respectively.
%$\text{vec}(\cdot)$ and $\text{diag}(\cdot)$ denote the vectorization operation and diagonalization operation, respectively.
% $\vec{a}\times\vec{b}$ denotes the cross product of vectors $\vec{a}$ and $\vec{b}$.

\section{System model}\label{sec:sys_model}

\begin{figure}[htbp]
    \centering
    \includegraphics[width=\linewidth]{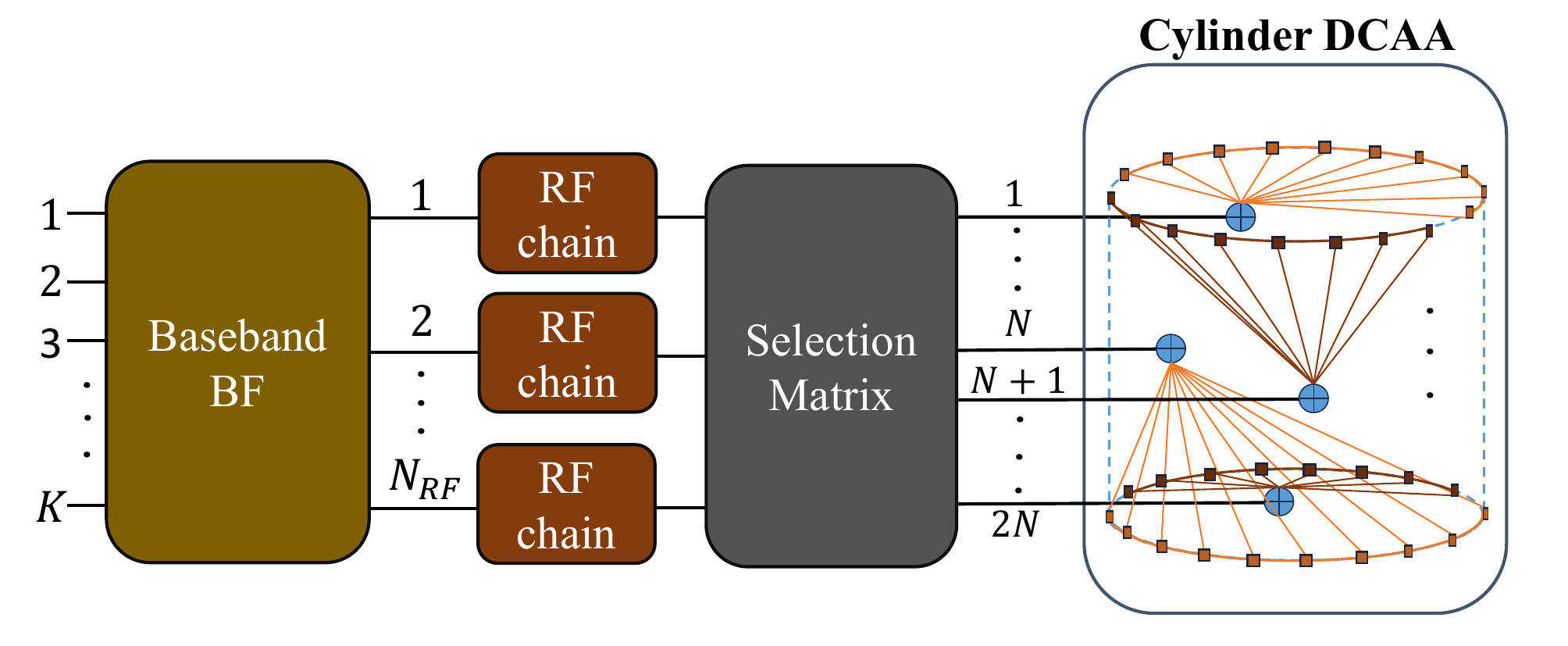}
    \caption{The overall scheme of communication system with cylinder DCAA. $2N$ sub-arrays are selectively connected to $N_\text{RF}$ RF chains through the selection matrix. }
    \label{fig:cylinder_DCAA}
\end{figure}

As shown in Fig. \ref{fig:cylinder_DCAA}, we consider a wireless communication system with $N_\text{RF}$ RF chains supporting $K$ users, where $K\leq N_\text{RF}$. 
HBF is a classic architecture to meet such a requirement by combining low-dimensional baseband digital beamforming with high-dimensional analog beamforming.
However, for mmWave or THz XL-MIMO systems for 6G networks, the required large number of high-precision front-end RF components are expensive and hard to fabricate, which inevitably increases hardware costs and therefore hinders the practical implementation of XL-MIMO. 
To address this problem, a recently proposed RAA structure reduces hardware costs while enhancing wireless performance by achieving uniform spatial resolution and satisfying beamforming gain through the innovative ray-like array structure \cite{RAA_long}. 
However, as illustrated in Fig. \ref{fig:signal_blockage}(a), RAA may suffer from signal blockage issue with multiple differently orientated sULAs in the same plane. 

To address this problem, we propose a novel cylinder DCAA-based wireless communication system, as shown in Fig. \ref{fig:signal_blockage}(b) and Fig. \ref{fig:cylinder_DCAA}. 
The proposed cylinder DCAA has $N$ vertically stacked sUCAs and each sUCA is partitioned into two complementary directly-connected sub-arrays. 
To adaptively connect $2N$ sub-arrays to $N_{\text{RF}}$ RF chains for further digital baseband processing, a \emph{selection matrix} denoted by $\mathbf{S}\in\{0,1\}^{N_{\text{RF}}\times2N}$ which satisfies $\Vert[\mathbf{S}]_{i,:}\Vert=1$ and $\Vert[\mathbf{S}]_{:,n}\Vert\leq1$, $1\leq i \leq N_{\text{RF}}$ and $1\leq n\leq 2N$ is introduced. 
Through the selection matrix, the communication system with cylinder DCAA can serve $K$ users with the number of RF chains being $N_\text{RF}\ll N$ without analog beamforming, thus decreasing hardware costs and complexity.  
In the following section, we will discuss the characteristics of sUCA, which is the basic component of cylinder DCAA.

\section{Simple uniform circular array}\label{sec:sUCA}

\begin{figure}[!t]
	\centering
	\includegraphics[width=\linewidth]{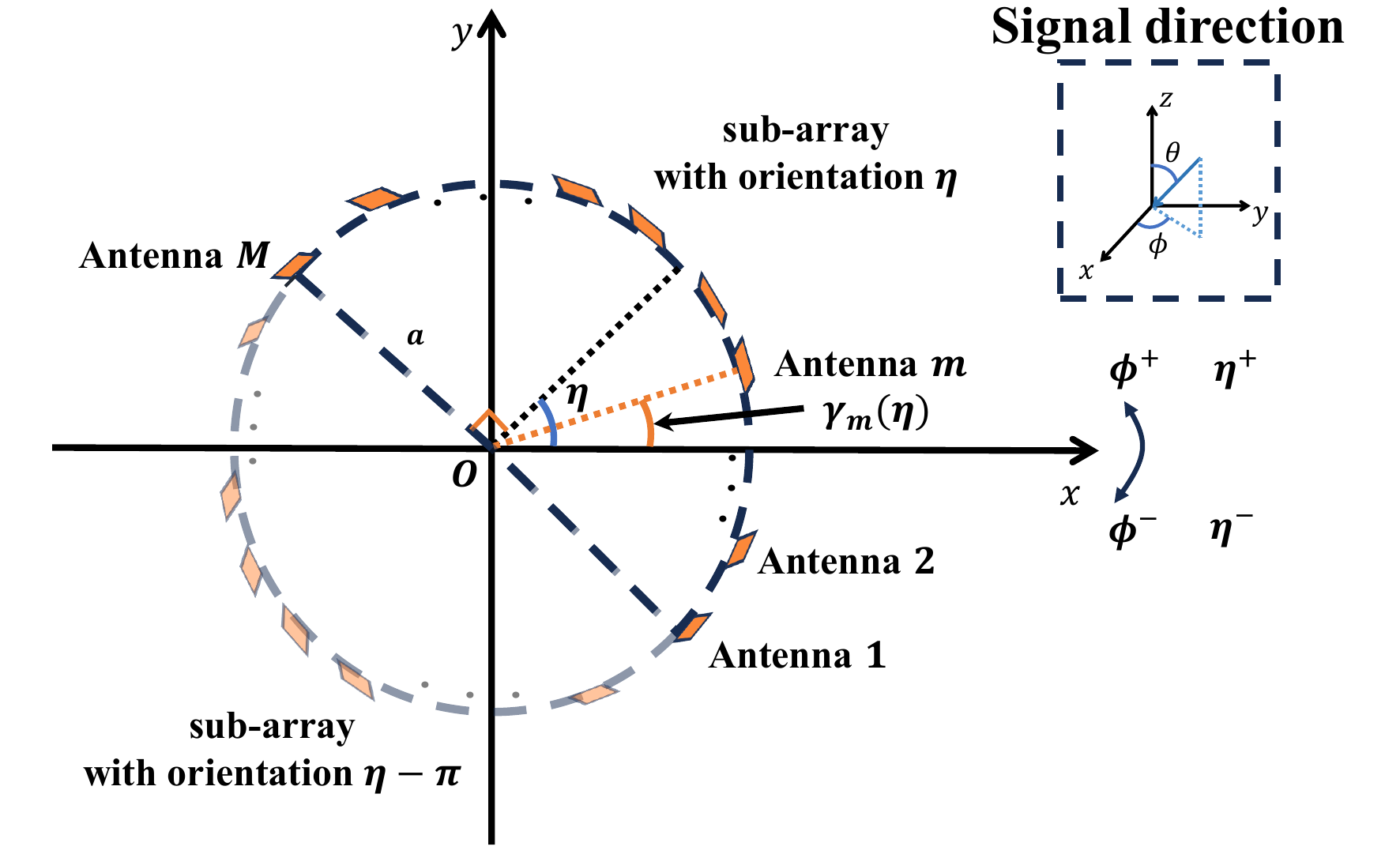}
	\caption{An illustration of sUCA with $2M$ antenna elements partitioned into two sub-arrays, where $M$ antenna elements are arranged into a uniform semi-circular architecture with orientation angle $\eta$ or $\eta-\pi$ relative to the positive $x$ axis and the $m\text{th}$ antenna element in the sub-array has the orientation angle $\gamma_m(\eta)$ or $\gamma_m(\eta-\pi)$, respectively.}
	\label{semi}
\end{figure}

% In this section, we present the geometric structure of sUCA and derive its response pattern. 
As illustrated in Fig. \ref{semi}, a sUCA consists of $2M$ uniformly spaced directional antenna elements along a circular ring of radius $a$ on the $xOy$ plane. 
The sUCA is equally partitioned into two sub-arrays each consisting of $M$ antenna elements, where the orientation angle $\eta$ of a sub-array is defined as the orientation angle of the central antenna element relative to the reference positive $x$ axis, and the $m\text{th}$ element in the sub-array with orientation $\eta$ has the orientation angle $\gamma_m(\eta)$, $m=1, 2, \cdots, M$. 
Since the response pattern of sUCA is the combination of response patterns of two complementary sub-arrays, in the following part, we will focus on the response pattern of a semi-circular-configured sub-array. 
The array's radius $a$ is given by:

\begin{equation}
	a=\frac{(M-1)\lambda}{2\pi}\approx\frac{M\lambda}{2\pi},\qquad M\gg1,
	\label{radius}
\end{equation}
where $\lambda$ is the wavelength. 
The orientation angle of the $m\text{th}$ antenna element $\gamma_m(\eta)$ relative to the positive $x$ axis is given by:
\begin{equation}
	\gamma_m(\eta)=\eta-\frac{\pi}{2}+\frac{\pi}{M-1}(m-1).
	\label{equ:thetam}
\end{equation}
Thus, the three dimensional location vector of the $m\text{th}$ antenna element is given by:
\begin{equation}
	\mathbf{v}_m(\eta)=[a\cos\gamma_m(\eta), a\sin\gamma_m(\eta), 0]^T.
\end{equation} 
The uniform plane wave (UPW) signal that comes from azimuth angle $\phi$ and zenith angle $\theta$ is considered. 
Its direction vector is given by $\mathbf{\textbf{k}}=[\cos\phi\sin\theta, \sin\phi\sin\theta, \cos\theta]^T$. The projection of $\mathbf{v}_m(\eta)$ on $\mathbf{\textbf{k}}$ is given by:
\begin{equation}
	\begin{aligned}\mathbf{v}_m^T(\eta)\mathbf{\textbf{k}}&=a\sin\theta(\cos(\gamma_m(\eta))\cos\phi+\sin(\gamma_m(\eta))\sin\phi)\\
	&=a\sin\theta\cos(\phi-\gamma_m(\eta)).
	\end{aligned}
\end{equation}
Thus, the array response vector of a sub-array with orientation angle $\eta$ for a UPW with $(\phi, \theta)$ is given by:
\begin{equation}
	\begin{aligned}
	\mathbf{a}(\eta,\phi,\theta)=&\left[ e^{-j\frac{2\pi}{\lambda}a\sin\theta\cos(\xi_1(\eta))}, e^{-j\frac{2\pi}{\lambda}a\sin\theta\cos(\xi_2(\eta))}, \right. \\ 
	&\left.\cdots ,e^{-j\frac{2\pi}{\lambda}a\sin\theta\cos(\xi_M(\eta))} \right]^T\odot \mathbf{b}(\eta,\phi,\theta),
	\end{aligned}
	\label{equ:response_vector}
\end{equation}
where $\mathbf{b}(\eta,\phi,\theta)$ represents the element pattern vector, which is given by:
\begin{equation}
	\mathbf{b}(\eta,\phi,\theta)=\left[\sqrt{G(\xi_1(\eta), \psi)},\cdots,\sqrt{G(\xi_M(\eta), \psi)}\right]^T,
	\label{equ:opt_b}
\end{equation}
where $G(\cdot)$ is the antenna element radiation pattern with $\xi_m(\eta) = \phi-\gamma_m(\eta)$ and $\psi = \theta - \frac{\pi}{2}$. 

\begin{figure}
    \centering
    \includegraphics[width=\linewidth]{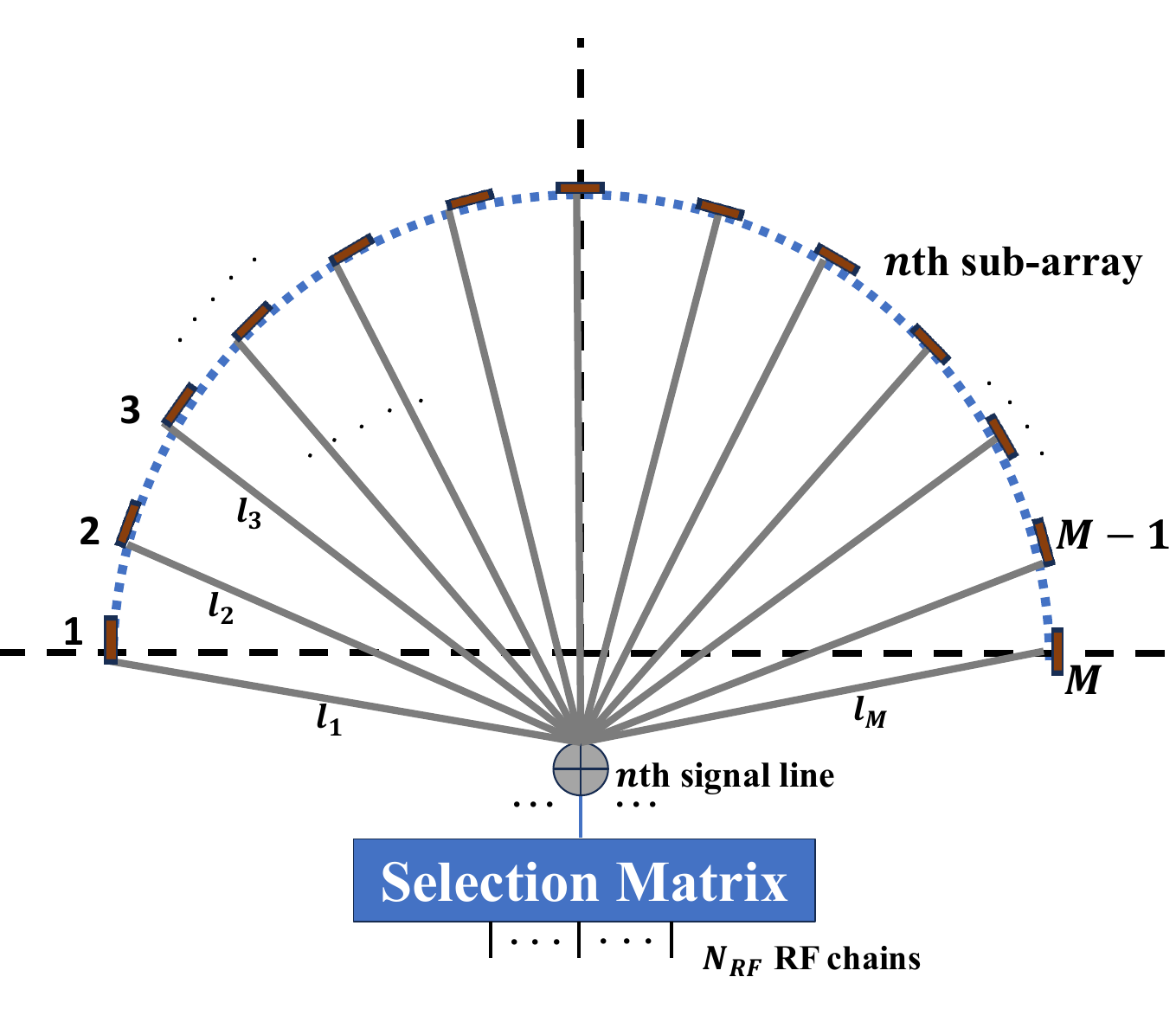}
    \caption{The variable delay line technology. $M$ antenna elements in a sub-array have different radiation phases based on carefully designed variable delay line. }
    \label{fig:feeding}
\end{figure}

In order to align the sub-array's main lobe with its physical orientation angle $\eta$, a variable length antenna element delay line technology is introduced \cite{ref:variable_length} to shift each antenna element's radiation phase. 
Since variable delay line technology employs passive delay lines with predefined delay values set via simple jumper configurations on the printed circuit board (PCB), the implementation will not introduce significant additional hardware costs and is relatively straightforward in terms of fabrication. 
As illustrated in Fig. \ref{fig:feeding}, the delay line's length of the $m\text{th}$ antenna element in a sub-array is given by $l_{m}$. 
Instead of fixed length antenna feeding where each antenna element has equal length delay line, i.e., $l_1=l_2=\cdots=l_M$ and the adaptive phase shift is achieved through the phase shifter of each antenna element, the variable-length delay line in Fig. \ref{fig:feeding} will introduce different radiation phases for every antenna element. 
By carefully designing the length of every element's delay line, we can align the main lobe of the sub-array to the desired direction. Array factor $\text{AF}(\eta,\phi,\theta)$ of the sub-array with orientation $\eta$ is given by:
\begin{equation}
\begin{aligned}	
	&\text{AF}(\eta,\phi,\theta)\\
	&=\mathbf{a}^T(\eta,\phi,\theta)\boldsymbol{\delta}(\eta,\theta)\\
	&=\sum_{m=1}^{M}\sqrt{G(\xi_m(\eta), \psi)}e^{-j(\frac{2\pi}{\lambda}a\sin\theta\cos\xi_m(\eta)-\varDelta\varphi_m(\eta,\theta))},
\end{aligned}
\label{equ:AF_origin}
\end{equation}
where $\boldsymbol{\delta}(\eta,\theta)$ is the phase shift vector achieved by the delay line, which is given by:
\begin{equation}
	\boldsymbol{\delta}(\eta,\theta)=\left[e^{j\varDelta\varphi_1(\eta,\theta)},e^{j\varDelta\varphi_2(\eta,\theta)},\cdots,  e^{j\varDelta\varphi_M(\eta,\theta)}\right]^T,
	\label{phasev}
\end{equation}
with $\varDelta\varphi_m(\eta,\theta)$ being the phase shift of the $m\text{th}$ antenna element introduced by the delay line. 
To simplify calculation, every antenna element's delay micro-strip line is considered as an ideal transmission line, thus the phase shift of the $m\text{th}$ antenna element in the sub-array with orientation $\eta$ is given by:
\begin{equation}
	\varDelta\varphi_m(\eta,\theta)=-\frac{2\pi}{\lambda}l_{m}(\eta,\theta).
	\label{phase}
\end{equation}
Similar to the RAA architecture \cite{RAA}, we hope the main lobe of the sub-array with orientation $\eta$ align with the direction $\phi=\eta$. Thus, the phase shift of the $m\text{th}$ antenna element $\varDelta\varphi_m(\eta,\theta)$ is given by:
\begin{equation}
	\begin{aligned}
	\varDelta\varphi_m(\eta,\theta)&=\frac{2\pi}{\lambda}a\sin\theta\cos\left(\eta-\gamma_m(\eta)\right)+2k\pi\\
	&=\frac{2\pi}{\lambda}a\sin\theta\sin\left(\frac{\pi}{M-1}(m-1)\right)+2k\pi, 
	\end{aligned}
	\label{phaseshift}
\end{equation}
where $k=1,2,\cdots$. From \eqref{phase} and \eqref{phaseshift}, the delay line's length of the $m\text{th}$ antenna element is given by:
%\begin{equation}
%	l_m^\eta=l_m=a\sin\theta\sin\left(\frac{\pi}{M-1}(m-1)\right)\quad(1\leq m\leq M),
%	\label{feedinglength}
%\end{equation}
\begin{equation}
	\begin{aligned}
		&l_m(\eta,\theta)=\underset{k_m}{\min}\left[-a\sin\theta\sin\left(\frac{\pi}{M-1}(m-1)\right)+k_m\lambda\right]\\
		&\text{s.t.}
		\begin{cases}
			l_m(\eta,\theta)>0\\
			k_m=1,2,\cdots,
		\end{cases}
	\end{aligned}
	\label{feedinglength}
\end{equation}
where $k_m$ corresponds the integer $k$ in \eqref{phaseshift} for the $m\text{th}$ antenna element. 
Based on the fact that the link distance is much greater than the height difference between the transmitter and receiver, we only consider the signal comes around the $xOy$ plane in the parameters design process, i.e., $\theta\approx\frac{\pi}{2}$.
It is obvious that the $m\text{th}$ and the $(M+1-m)\text{th}$ antenna element have the equal minimum length of the delay line and $l_m(\eta,\theta)=l_m$ is independent of the orientation angle $\eta$, such that the variable-length delay line design has good compliance with the principle of symmetry. 
% The approximation $\theta\approx\frac{\pi}{2}$ is reasonable when the link distance is much greater than the height difference between the transmitter and receiver. 
Thus, the length of minimum delay line in \eqref{feedinglength} is reduced to:
\begin{equation}
		l_m=-a\sin\left(\frac{\pi}{M-1}(m-1)\right)+k_m\lambda,
	\label{equ:reduced_feeding}
\end{equation}
where the integer $k_m$ that keeps the non-negativity of $l_m$ is formulated by:
\begin{equation}
    k_m=\left\lceil
    \frac{a}{\lambda}\sin\left(\frac{\pi}{M-1}(m-1)\right)
    \right\rceil.
    \label{equ:k_m}
\end{equation}
Based on \eqref{equ:reduced_feeding}, the phase shift $\varDelta\varphi_m(\eta,\theta)=\varDelta\varphi_m$ in \eqref{phaseshift} and the phase shift vector $\boldsymbol{\delta}(\eta,\theta)=\boldsymbol{\delta}$ in \eqref{phasev} are independent of physical orientation of the sub-array. 

% Fig. \ref{radiation} illustrates the response pattern of a directly-connected sub-array given in Fig. \ref{semi}, with $M=32$ under central frequency $f_c=47.2 \text{GHz}$. The antenna element's radiation pattern $G(\xi,\psi)$ adopts the 3D 3GPP model \cite{3GPP} whose representation in dB is given by: 
% \begin{subequations}\label{antenna}
% 	\begin{equation}
% 		\begin{aligned}
% 			&G_\text{dB}(\xi_m,\psi)=\\
% 			&-\min\left[-\left(A_{\text{dB}}\left(\xi_m ,\psi = 0\right)+A_{\text{dB}}\left(\xi_m=0, \psi\right)\right), A_{\text{max}}\right]
% 		\end{aligned}
% 	\end{equation}
% 	\begin{equation}
% 		\begin{cases}
% 			&A_{\text{dB}}\left(\xi_m ,\psi = 0\right)=-\min\left[12\left(\frac{\xi_m}{\xi_{\text{3dB}}}\right)^2,A_{\text{max}}\right]\\[8pt]
			
% 			&A_{\text{dB}}\left(\xi_m=0, \psi\right)=-\min\left[12\left(\frac{\psi}{\psi_{\text{3dB}}}\right)^2, \text{SLV}_v\right]\\[8pt]
			
% 			&A_{\text{max}}=\text{SLV}_v=30\text{dB}\\[8pt]
% 			&\xi_{\text{3dB}}=65^{\circ}\\[8pt]
% 			&\psi_{\text{3dB}}=65^{\circ},\\[8pt]
% 		\end{cases}
% 	\end{equation}
% \end{subequations}
% where $G_\text{dB}(\xi_m,\psi)=10\log_{10}(G(\xi_m,\psi))$ is the antenna gain expressed in dB. 

% \begin{figure}[!t]
% 	\centering
% 	\subfloat[Fixed length antenna delay line.]{%
% 		\includegraphics[width=\linewidth]{conventional.pdf}
% 		\label{conventional}
% 	}
% 	\hfil
% 	\subfloat[Variable length antenna delay line.]{%
% 		\includegraphics[width=\linewidth]{variable.pdf}
% 		\label{variable}
% 	}
% 	\caption{Two types of antenna delay line. $M$ antenna elements in a sub-array have different radiation phase based on carefully designed variable length of the feeding line in (b).} 
% 	\label{feeding}
% \end{figure}

\begin{figure}[!t]
	\centering
	\includegraphics[width=\linewidth]{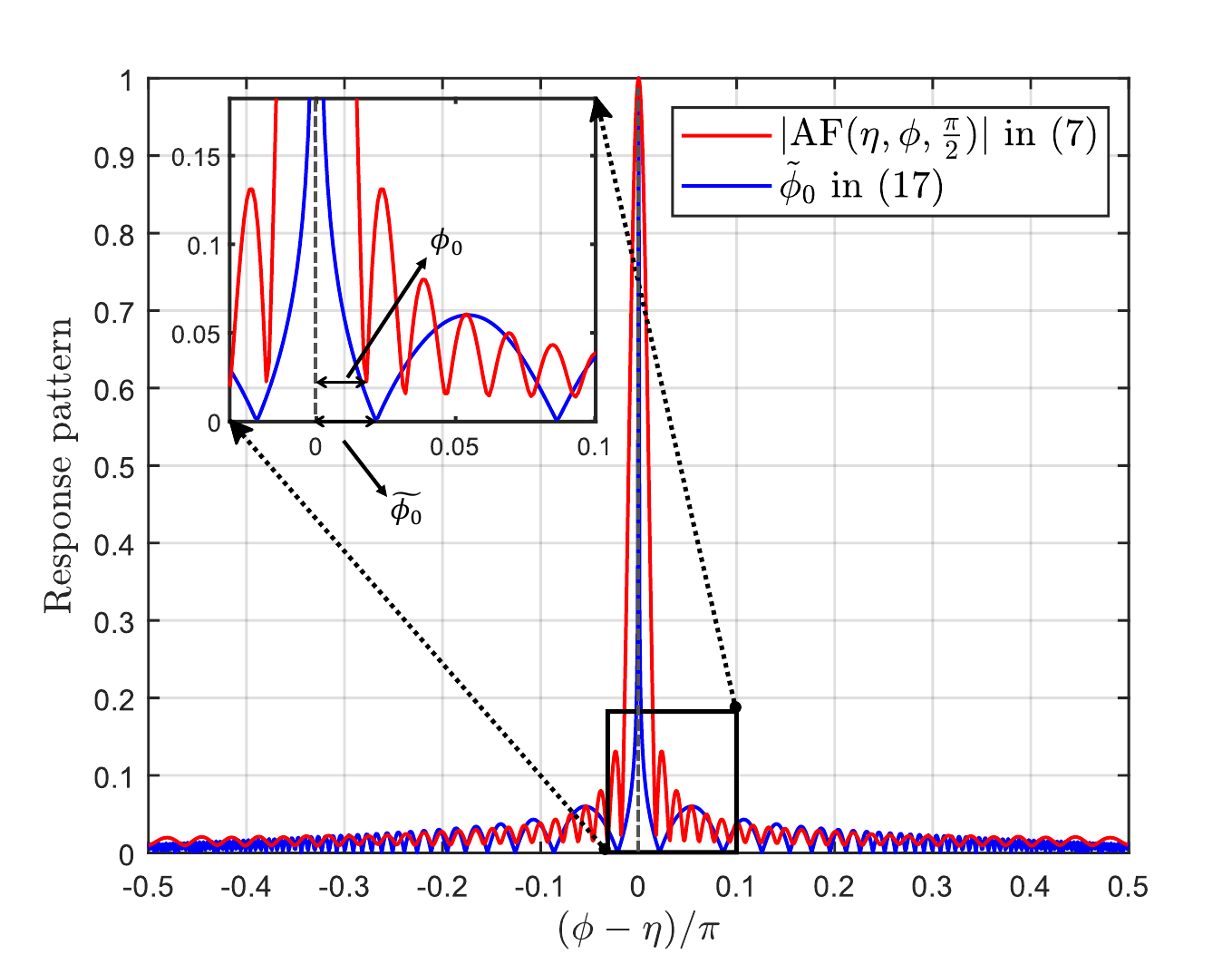}
	\caption{Comparison of theoretical and approximated valley point of the uniformed $\vert\text{AF}(\eta,\phi,\pi/2)\vert$ in \eqref{equ:AF_origin} and \eqref{equ:valleypoint} with $M=64$ under carrier frequency $f_c=47.2\text{GHz}$. 
	Results show that the approximated $\tilde{\phi}_0$ is very close to the theoretical $\phi_0$.} 
	\label{fig:appro_vs_af}
\end{figure}

With the variable length antenna delay line technology, array factor of the sub-array with orientation $\eta$ is further expressed by:
\begin{equation}
	\begin{aligned}
		&\text{AF}(\eta,\phi,\theta)\\
		&=\sum_{m=1}^{M}\sqrt{G(\xi_m(\eta),\psi)}e^{-j\frac{2\pi}{\lambda}a\sin\theta\left(\cos\xi_m(\eta)-\cos(\eta-\gamma_m(\eta))\right)}\\
		&=\sum_{m=1}^{M}\sqrt{G(\xi_m(\eta),\psi)}e^{j\frac{4\pi}{\lambda}a\sin\theta\sin\left(\frac{\phi+\eta-2\gamma_m(\eta)}{2}\right)\sin\left(\frac{\phi-\eta}{2}\right)}\\
		&=\sum_{m=1}^{M}\sqrt{G(\xi_m(\eta),\psi)}e^{j\frac{4\pi}{\lambda}a\sin\theta\cos\left(\frac{\phi-\eta}{2}-\frac{\pi}{M-1}(m-1)\right)\sin\left(\frac{\phi-\eta}{2}\right)}
	\end{aligned}
	\label{AF_temp}
\end{equation}
With Jacobi-Anger expansion, the response can be further expressed by:
	\begin{equation}
		\begin{aligned}
			&\text{AF}(\eta,\phi,\theta)\\
            &=\sum_{m=1}^{M}\sqrt{G(\xi_m(\eta), \psi)}\times\\
			&\sum_{n=-\infty}^{+\infty}j^nJ_n\left(\frac{4\pi}{\lambda}a\sin\theta\sin\frac{\phi-\eta}{2}\right)e^{jn\frac{\phi-\eta}{2}}e^{-jn\frac{\pi(m-1)}{M-1}}\\
			&=\sum_{n=-\infty}^{+\infty}j^ne^{jn\frac{\phi-\eta}{2}}J_n\left(\frac{4\pi}{\lambda}a\sin\theta\sin\frac{\phi-\eta}{2}\right)\times\\
			&\sum_{m=1}^{M}\sqrt{G(\xi_m(\eta), \psi)}e^{-jn\frac{\pi(m-1)}{M-1}}\\
			&=\sum_{n=-\infty}^{+\infty}j^ne^{jn\frac{\phi-\eta}{2}}J_n\left(\frac{4\pi}{\lambda}a\sin\theta\sin\frac{\phi-\eta}{2}\right)\text{S}_n(\eta,\phi,\theta), 
		\end{aligned}
		\label{AF}
	\end{equation}
where $J_n(\cdot)$ refers to the Bessel function of the first kind and $\text{S}_n(\eta,\phi)$ is given by:
\begin{equation}
	\text{S}_n(\eta,\phi,\theta)=\sum_{m=1}^{M}\sqrt{G(\xi_m(\eta), \psi)} e^{-jn\frac{\pi}{M-1}(m-1)}.
	\label{equ:Sn}
\end{equation}
Based on the characteristics of the Bessel function of the first kind, the position of the first valley point of $\vert\text{AF}(\eta,\phi,\theta)\vert$ mainly depends on the zeroth-order Bessel function $J_0(\cdot)$. 
To get an approximate but explicit expression of the distance $\phi_0=\vert\phi_\text{valley}-\eta\vert$ between the first valley point $\phi_\text{valley}$ of $\vert\text{AF}(\eta,\phi,\theta)\vert$ relative to azimuth angle $\phi$ and the main lobe's direction $\eta$, the approximated value $\tilde{\phi_0}$ is modeled as the first zero point of $\left.J_0\left(\frac{4\pi}{\lambda}a\sin\theta\sin\frac{\phi}{2}\right)\right\vert_{\theta=\frac{\pi}{2}}$. Thus, the approximation $\tilde{\phi}_0$ is given by:
%i.e.,  $\vert\textbf{AF}^\eta(\phi,\theta)\vert_{\min}=\vert\textbf{AF}^\eta(\phi_\text{valley},\theta)\vert$, and the main lobe direction $\eta$, we may express it by the first zero point of the 0 order Bessel function:
\begin{equation}
	\tilde{\phi}_0=2\arcsin\left(\frac{3.83\lambda}{4\pi a}\right)=2\arcsin\left(\frac{3.83}{2M}\right).
	\label{equ:valleypoint}
\end{equation}
With $M\gg1$ in XL-MIMO system, \eqref{equ:valleypoint} is simplified to $\tilde{\phi_0}\approx\frac{3.83}{M}$. 
% In the following section, we will prove that this assumption is suitable for mmWave communication in \emph{Proposition} \ref{prop:farfield}. 
As illustrated in Fig. \ref{fig:appro_vs_af}, $\tilde{\phi_0}$ is very close to the exact $\phi_0$ in \eqref{equ:AF_origin}, where antenna elements' radiation pattern $G(\xi,\psi)$ adopts the 3D 3GPP technical report \cite{3GPP} whose representation in dB is given by: 
\begin{subequations}\label{antenna}
	\begin{equation}
		\begin{aligned}
			&G_\text{dB}(\xi_m,\psi)=\\
			&-\min\left\{-\left(A_{\text{dB}}\left(\xi_m ,\psi = 0\right)+A_{\text{dB}}\left(\xi_m=0, \psi\right)\right), 30 \text{ dB}\right\}
		\end{aligned}
	\end{equation}
	\begin{equation}
		\begin{cases}
			&A_{\text{dB}}\left(\xi_m ,\psi = 0\right)=-\min\left[12\left(\frac{\xi_m}{65^{\circ}}\right)^2,30 \text{ dB}\right]\\[8pt]
			
			&A_{\text{dB}}\left(\xi_m=0, \psi\right)=-\min\left[12\left(\frac{\psi}{65^{\circ}}\right)^2, 30 \text{ dB}\right]\\[8pt]
		\end{cases}
	\end{equation}
\end{subequations}
with $G_\text{dB}(\xi_m,\psi)=10\log_{10}(G(\xi_m,\psi))$ being the antenna gain expressed in dB. 

\begin{figure}[!t]
	\centering
	\includegraphics[width=\linewidth]{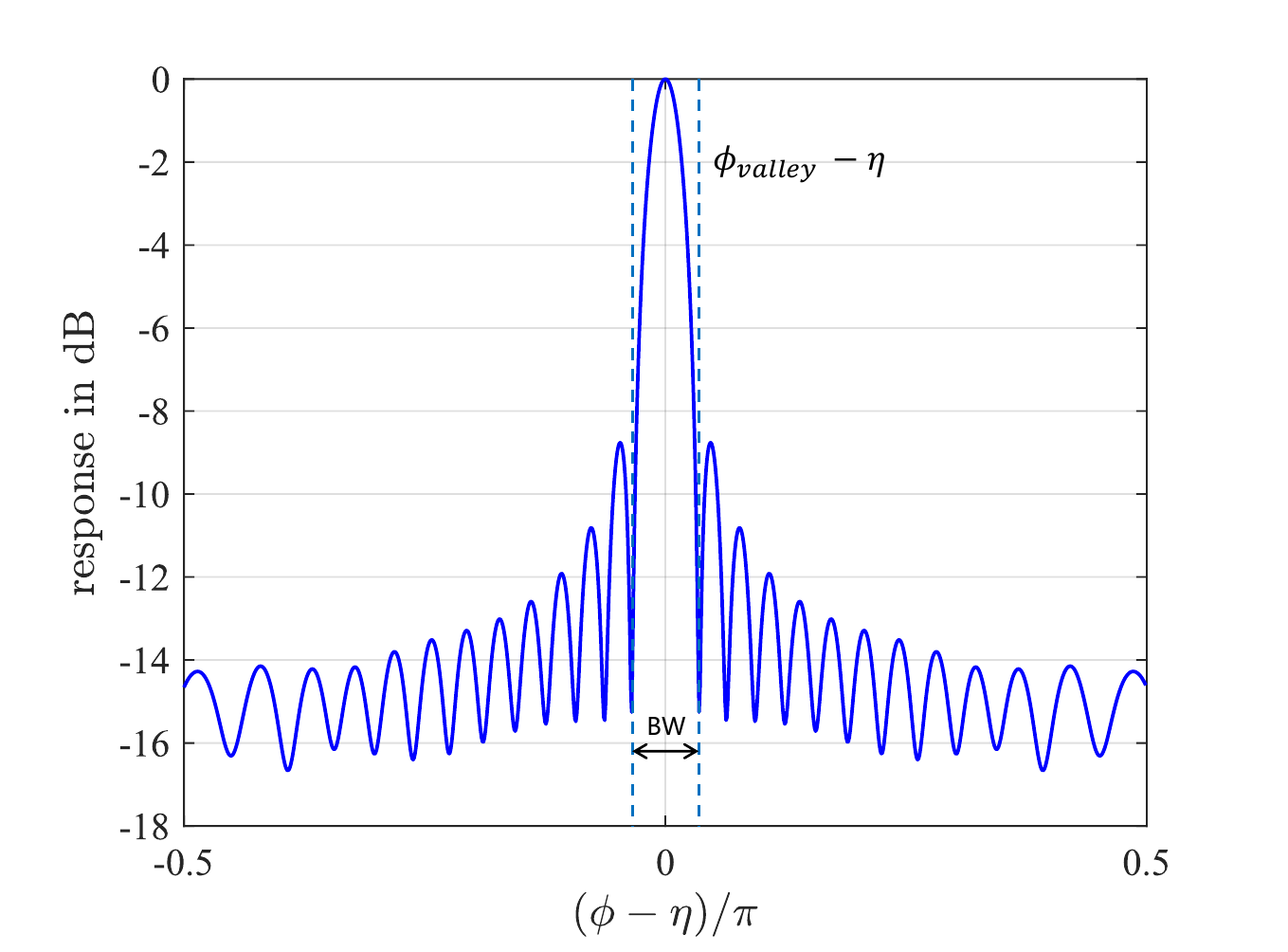}
	\caption{Response pattern $\vert\text{AF}(\eta,\phi,\pi/2)\vert$ of the sub-array with orientation $\eta$ and $M=32$ in \eqref{equ:AF_origin} with variable length delay line.}
	\label{radiation}
\end{figure}

Fig. \ref{radiation} illustrates the response pattern $\vert\text{AF}(\eta,\phi,\pi/2)\vert$ given in \eqref{equ:AF_origin} of a directly-connected sub-array given in Fig. \ref{semi}, with $M=32$ under central frequency $f_c=47.2 \text{GHz}$. 
The results reveal that the proposed directly-connected sub-array aligns its main beam lobe to its physical orientation, implying its advance performance in spatial resolution and interference suppression. 
Based on \eqref{equ:AF_origin} and \eqref{AF}, we can derive \emph{Theorem} \ref{the:AF} that illustrates the characteristics of sub-array's response pattern. 
\begin{theorem}
	\label{the:AF}
	The maximum value of array factor $\vert\text{AF}(\eta,\phi,\theta)\vert$ given in \eqref{AF} appears at angle set $(\phi,\theta)=(\eta,\frac{\pi}{2})$. The beam width of $\vert\text{AF}(\eta,\phi,\theta)\vert$ is given by $\text{BW}\approx2\tilde{\phi_0}=4\arcsin\left(\frac{3.83}{2M}\right)$.
\end{theorem}
\begin{IEEEproof}
	Please refer to Appendix \ref{app:proof2}.
\end{IEEEproof}

\emph{Theorem} \ref{the:AF} reveals that sub-array can align the compact main beam lobe to the direction relative to its physical orientation $\eta$. 
For $M\gg1$, the beam width is approximated as $\text{BW}\approx\frac{7.66}{M}$, which decreases with the number of antenna elements $M$ in a sub-array. 
% which significantly improves the spatial resolution by concentrating power to a more compact main beam lobe for XL-MIMO with considerable antenna elements.  

% \section{system model for cylinder DCAA-based wireless communication}

% \begin{figure}[!t]
%     \centering
%     \includegraphics[width=\linewidth]{cylinder_DCAA.pdf}
%     \caption{The overall scheme of communication system with cylinder DCAA. $2N$ sub-arrays are selectively connected to $N_\text{RF}$ RF chains through the selection matrix. }
%     \label{fig:cylinder_DCAA}
% \end{figure}

% As shown in Fig. \ref{fig:cylinder_DCAA}, we consider a wireless communication system with $N_\text{RF}$ chains supporting $K$ users, where $K<N_\text{RF}$. 
% The HBF is a classic architecture to meet such a requirement by combining low-dimensional baseband digital beamforming with high-dimensional analog beamforming.
% However, for XL-MIMO implemented in mmWave or THz frequency band for 6G network, number of high-precision front-end RF components are relative expensive and hard to fabricate. 

\section{Proposed Cylinder DCAA}\label{sec:DCAA}
In this section, based on the sUCA presented in Section \ref{sec:sUCA}, a novel cylinder DCAA architecture is proposed. 
% Detailed structure and parameter design are given as follows. 

\subsection{Cylinder DCAA basic structure}\label{sub_sec:define}

\begin{figure}[htbp]
	\centering
	\includegraphics[width=\linewidth]{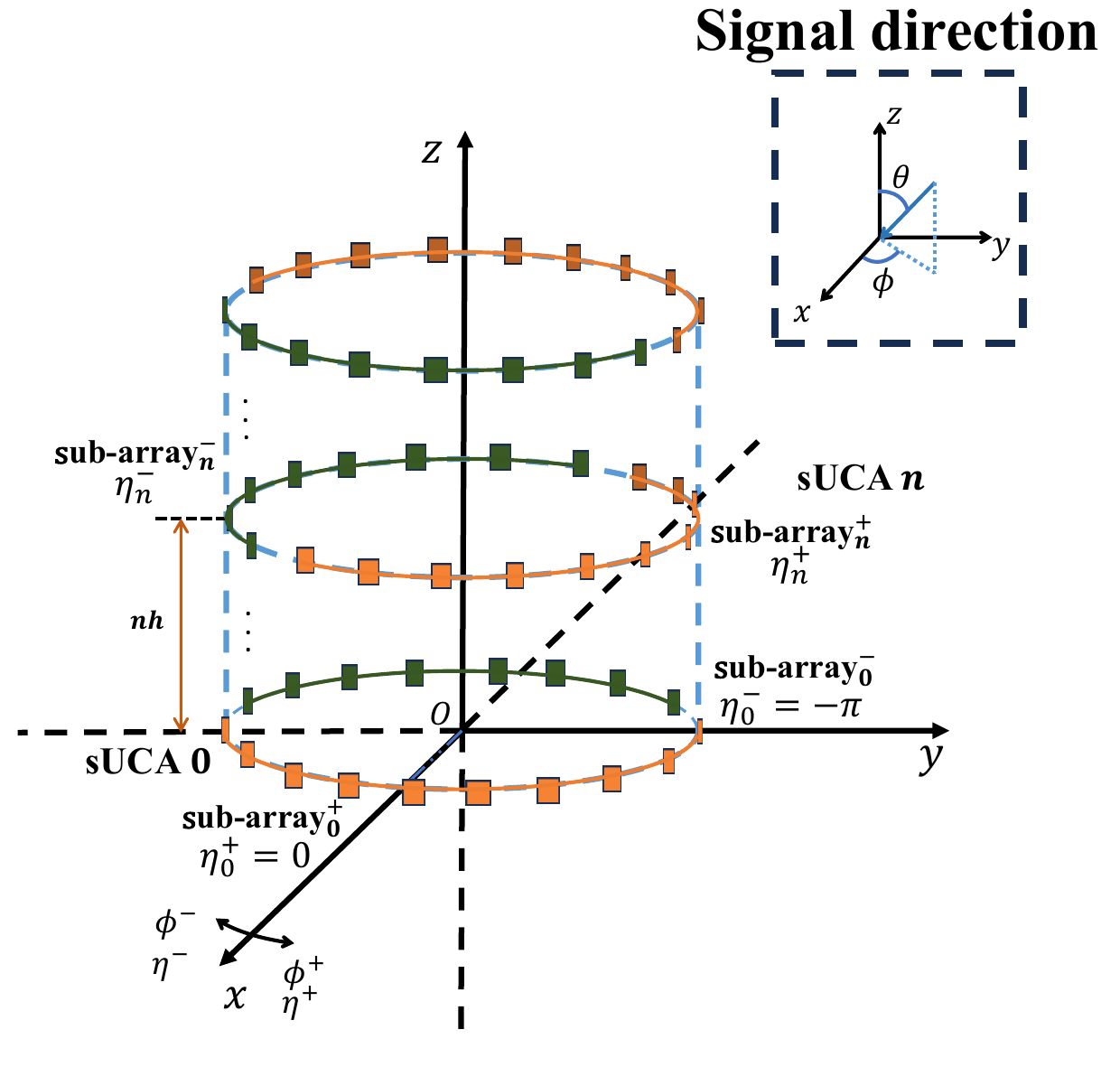}
	\caption{Cylinder DCAA with $N$ sUCAs with each of them consists of two complementary sub-arrays. 
	$\text{sub-array}_n^+$ and $\text{sub-array}_n^-$ has the orientation angle of $\eta_n^+$ and $\eta_n^-$ relative to the positive $x$ axis, respectively. 
	}
	\label{cylinder}
\end{figure}

As illustrated in Fig. \ref{cylinder}, a cylinder DCAA is composed of $2N$ sub-arrays arranged by $2N\times M$ antenna elements, where each sub-array has $M$ antenna elements, and every two sub-arrays that have complementary orientation angles, i.e., the difference of their orientation angles equals to $\pi$, are merged into a sUCA. 
Each sub-array has the antenna elements directly connected together through variable delay line technology without any digital or analog beamforming. 
Without loss of generality, a 3D Cartesian coordinate system is established, where the positive $x$ axis is chosen as the reference direction for azimuth angle $\phi$ and sub-array's orientation $\eta$. 

Since each sUCA is equally partitioned into two complementary sub-arrays, for sub-arrays in the $n\text{th}$ sUCA, we use $\text{sub-array}_n^+$ and $\text{sub-array}_n^-$ to denote the sub-array that has positive and negative orientation angle denoted by $\eta_n^+>0$ and $\eta_n^-<0$ relative to the reference direction, respectively. 
% For notational convenience, we use $\eta_n^+>0$ and $\eta_n^-<0$ to denote the central point's orientation angles relative to reference direction for $\text{sub-array}_n^+$ and $\text{sub-array}_n^-$, respectively.

Since $N$ sUCAs need to be arranged into a 3D layered structure, we define $h$ as the distance between vertically adjacent sUCAs. 
Without loss of generality, we choose the positive $z$ axis as the direction where a sUCA is stacked. 
Hence, the height of cylinder DCAA that has $N$ sUCAs is given by $h_{\text{sum}}=(N-1)h$. 
Thus, a cylinder DCAA is characterized by the parameters $\{M,N,\{\eta_n^+\},\{\eta_n^-\},h_{\text{sum}}\}$. 

\subsection{Cylinder DCAA parameter design}\label{sub_sec:design}
To minimize interference between vertically adjacent sub-arrays, from the beamwidth $\text{BW}$ given in \emph{Theorem} \ref{the:AF}, their orientations are designed so that one sub-array's main lobe locates at the vertically adjacent sub-array's valley point, i.e., 
\begin{equation}
	\begin{aligned}
	&\eta_n^+=2n\arcsin\left(\frac{3.83}{2M}\right)\\
	&\eta_n^-=\eta_n^+-\pi=2n\arcsin\left(\frac{3.83}{2M}\right)-\pi.
	\end{aligned}
	\label{eta}
\end{equation}
%For design and calculation simplicity, we set a priori condition that  $\theta=\frac{\pi}{2}$ and $M\gg1$, thus orientation angle of the $\text{sub-array}_n^+$ is further given by $\eta_n^+=n\times\frac{3.83}{M}$.
Based on the fact that the link distance is much greater than the height difference between the transmitter and receiver XL-MIMO, we obtain that $M\gg1$ and $\theta=\frac{\pi}{2}$, thus orientation angle of the $\text{sub-array}_n^+$ is further given by $\eta_n^+=n\times\frac{3.83}{M}$. 

To cover the full-angle domain, the maximum orientation angle should satisfy  $\eta_{N-1}^+=\pi-\phi_0$. 
Thus, the number of sUCA that consists of two complementary sub-arrays is:
\begin{equation}
	N=\left\lfloor\frac{\pi M}{3.83}\right\rfloor.
	\label{N}
\end{equation}
Some typical $(M,N)$ sets are given by $(16,13), (32,26), (64,52)$ and $(128,104)$. For the $\text{sub-array}_n^+$, the response vector is obtained by substituting $\eta$ with $\eta_n^+$ in \eqref{equ:response_vector}:
\begin{equation}
	\begin{aligned}
		\mathbf{a}(\eta_n^+,\phi,\theta)=&\left[e^{-j\frac{2\pi}{\lambda}a\sin\theta\cos\xi_1(\eta_n^+)}, e^{-j\frac{2\pi}{\lambda}a\sin\theta\cos\xi_2(\eta_n^+)}\right. \\ 
		&\left. ,\cdots, e^{-j\frac{2\pi}{\lambda}a\sin\theta\cos\xi_M(\eta_n^+)} \right]^T\odot\mathbf{b}(\eta_n^+,\phi,\theta),
	\end{aligned}
	\label{equ:sca}
\end{equation}
where the antenna radiation pattern vector $\mathbf{b}(\eta_n^+, \phi,\theta)$ is obtained by substituting $\eta$ with $\eta_n^+$ in \eqref{equ:opt_b}:
\begin{equation}
	\mathbf{b}(\eta_n^+,\phi,\theta)=\left[\sqrt{G(\xi_1(\eta_n^+), \psi)},\cdots,\sqrt{G(\xi_M(\eta_n^+), \psi)}\right]^T.
	\label{equ:b}
\end{equation}
% where the relative arrival angle $\xi_m(\eta_n^+)=\phi-\gamma_m(\eta_n^+)$. %For the $\text{SCA}_n^-$, its response vector 
The response vector of $\text{sub-array}_n^-$ is obtained by substituting $\eta_n^+$ in \eqref{equ:sca} and \eqref{equ:b} with the complementary orientation angle $\eta_n^-=\eta_n^+-\pi$ in \eqref{eta}.

The $n\text{th}$ sUCA has the height of $nh$ relative to the $xOy$ plane, where we choose $h=\frac{\lambda}{2}$ to separate vertically adjacent sUCAs by half wavelength.
Hence, the total height of the cylinder DCAA is given by $h_{\text{sum}}=\frac{(N-1)\lambda}{2}$. 

% \begin{figure}[!t]
% 	\centering
% 	\includegraphics[width=\linewidth]{response2.pdf}
% 	\caption{Response pattern of cylinder DCAA, whose $M=16$, $N=13$ and the system works under central frequency $f_c=47.2\text{GHz}$. Cylinder DCAA achieves uniform spatial resolution with multiples sUCAs arranged with different physical orientations. }
% 	\label{BS_res}
% \end{figure}

\begin{figure}[!t]
	\centering
	\subfloat[Front view.]{%
		\includegraphics[width=\linewidth]{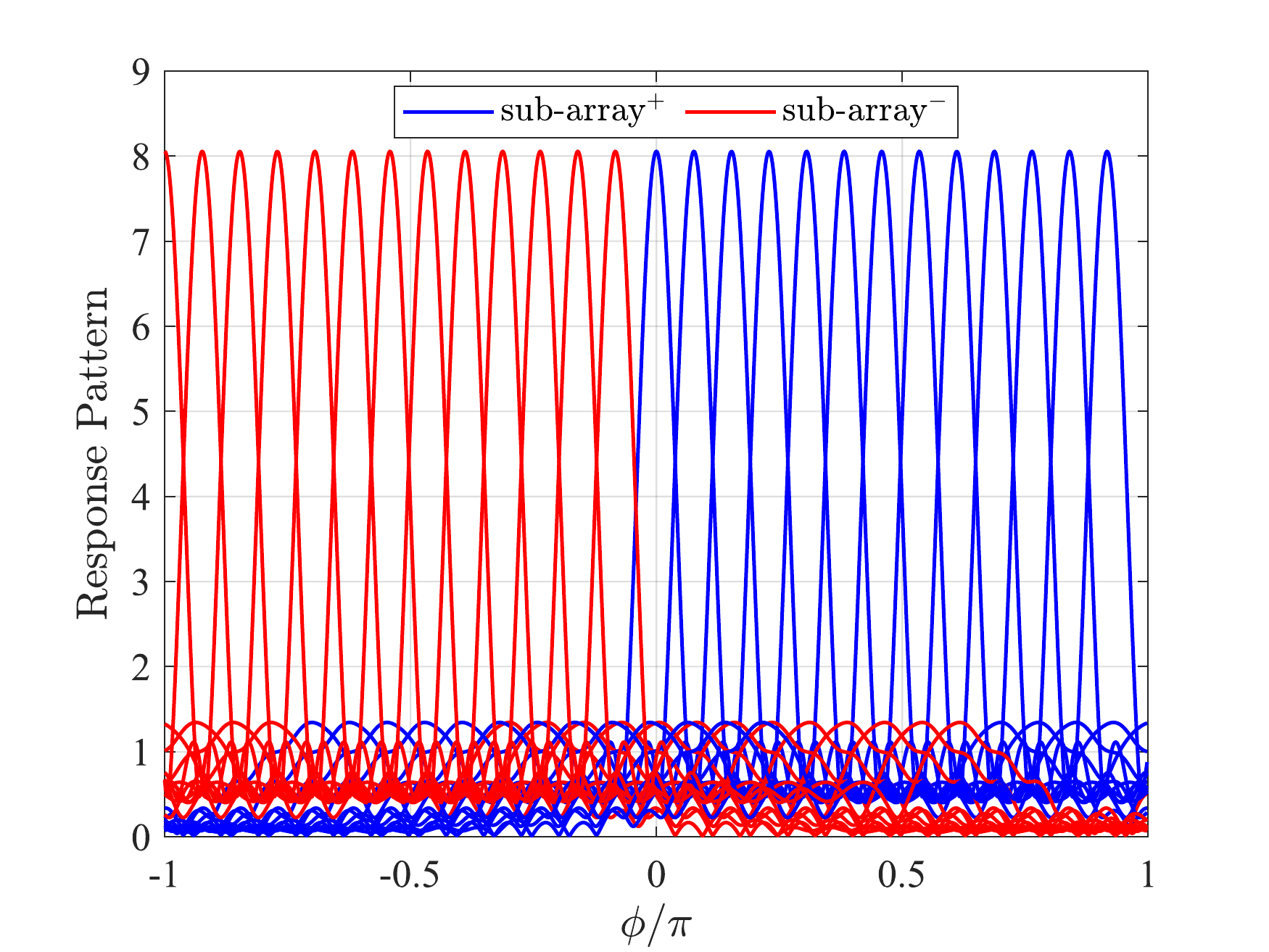}
		\label{fig:response_2}
	}
	\hfil
	\subfloat[Vertical view.]{%
		\includegraphics[width=\linewidth]{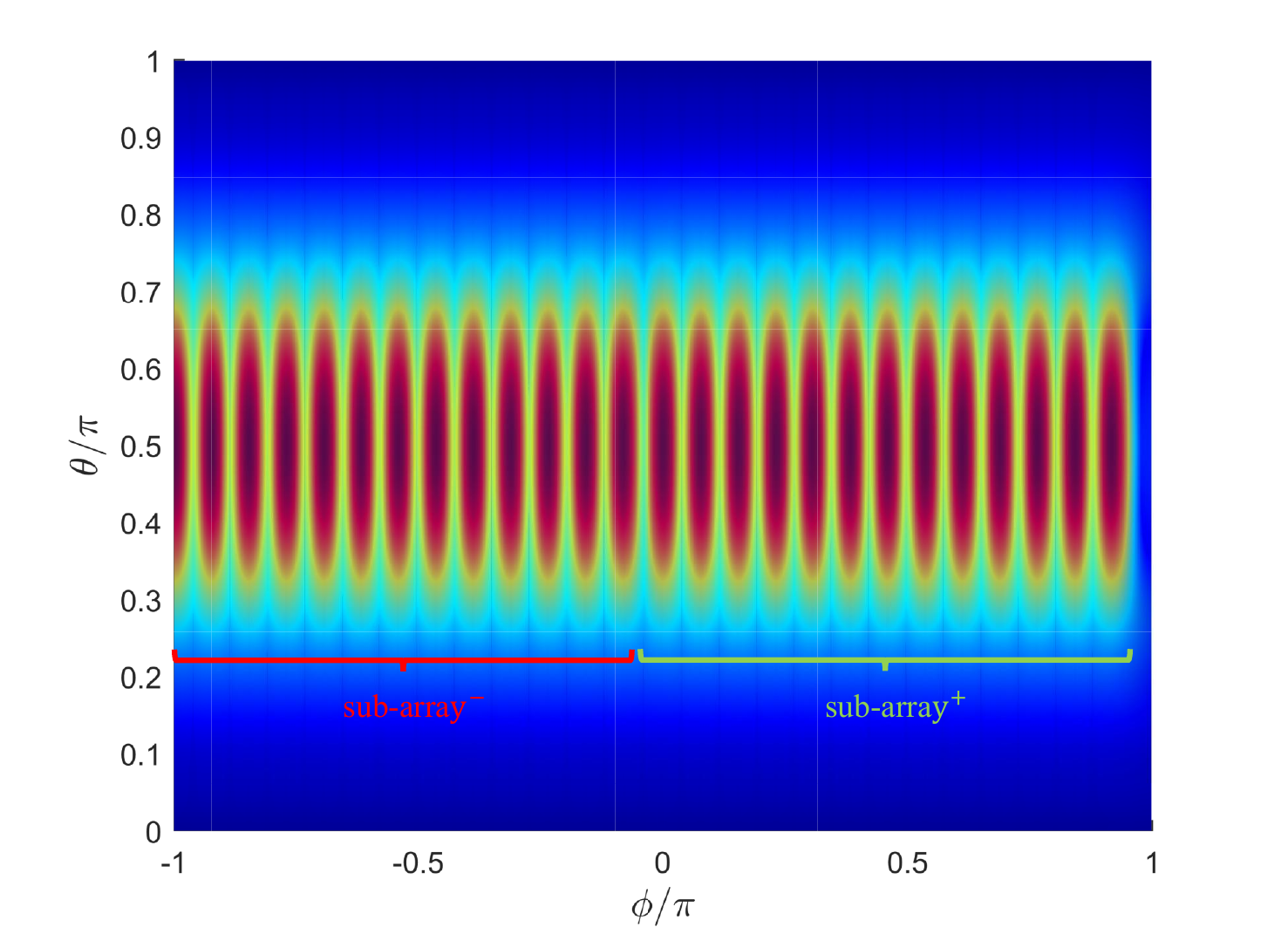}
		\label{fig:response_height}
	}
	\caption{Response pattern of cylinder DCAA with $M=16$, $N=13$. Cylinder DCAA achieves uniform spatial resolution with multiples sUCAs arranged with different physical orientations.} 
	\label{BS_res}
\end{figure}

The response pattern of cylinder DCAA with $M=16$ directional antenna elements and $N=13$ is shown in Fig. \ref{BS_res}, illustrating that the cylinder DCAA achieves uniform spatial resolution without relying on any front-end phase shifter, where $\text{sub-arrays}^+$ uniformly cover the positive angle domain $\phi\in[0,\pi)$ and $\text{sub-arrays}^-$ uniformly cover the negative angle domain $\phi\in[-\pi,0)$. 
The design of orientation angle $\eta_n^+$ and $\eta_n^-$ ensures that the sub-arrays will have as little interference as possible among vertically adjacent sub-arrays. 

% The former design and analysis of cylinder DCAA use \emph{a priori} approximation $\theta\approx\frac{\pi}{2}$, we derive \emph{Proposition} \ref{prop:farfield} to prove the effectivity of the approximation. 

% \begin{proposition}
% 	\label{prop:farfield}
% 	The far field UPW mmWave signal's zenith arrival angle can be approximated to $\theta\approx\frac{\pi}{2}$ for the design and analysis of cylinder DCAA with specific communication scenarios, such as in-door office and so on. 
% \end{proposition}

% \begin{IEEEproof}
% 	Please refer to Appendix \ref{app:proof1}.
% \end{IEEEproof}

% \begin{figure}[!t]
% 	\centering
% 	\includegraphics[width=\linewidth]{BS.pdf}
% 	\caption{Cylinder DCAA with $N$ sUCAs with each of them consists of two complementary sub-arrays. 
% 	$\text{sub-array}_n^+$ and $\text{sub-array}_n^-$ has the orientation angle of $\eta_n^+$ and $\eta_n^-$ relative to the positive $x$ axis, respectively. 
% 	}
% 	\label{cylinder}
% \end{figure}

\section{Wireless Communication with cylinder DCAA}\label{sec:signal_model}
In this section, we consider wireless communication systems with the proposed cylinder DCAA and the benchmark cell sectoring system based on ULA with HBF. 
\subsection{The proposed cylinder DCAA}
We first present the response matrix $\mathbf{A}(\phi,\theta)\in\mathbb{C}^{M\times2N}$ of communication system with cylinder DCAA illustrated in Fig. \ref{fig:cylinder_DCAA}. The response matrix is combined by 2 sub-matrices, i.e., $\mathbf{A}(\phi,\theta)=\left[\mathbf{A}^+(\phi,\theta),\mathbf{A}^-(\phi,\theta)\right]$, where $\mathbf{A}^+(\phi,\theta)\in\mathbb{C}^{M\times N}$ consists of the response vectors of $N$ sub-arrays with relatively positive orientation denoted by $\text{sub-arrays}^+$ and $\mathbf{A}^-(\phi,\theta)\in\mathbb{C}^{M\times N}$ consists of the response vectors of $N$ sub-arrays with relatively negative orientation denoted by $\text{sub-arrays}^-$. From \eqref{equ:sca}, the response matrix $\mathbf{A}(\phi,\theta)$ is given by:
\begin{equation}
\begin{aligned}
	&\mathbf{A}(\phi,\theta)\\
    &=
	\left[	\underbrace{
		\begin{bmatrix}
			\mathbf{a}(\eta_0^+),\cdots,\mathbf{a}(\eta_{N-1}^+)
		\end{bmatrix}
	}_{ \triangleq \mathbf{A}^+(\phi,\theta)},\right.
	\left.\underbrace{
		\begin{bmatrix}
			\mathbf{a}(\eta_0^-),\cdots,\mathbf{a}(\eta_{N-1}^-)
		\end{bmatrix}
	}_{ \triangleq \mathbf{A}^-(\phi,\theta)}
	\right],\\
\end{aligned}
	\label{equ:response_matrix}
\end{equation}
where $\mathbf{a}(\eta_n^+)$ stands for $\mathbf{a}(\eta_n^+, \phi, \theta)$ given in \eqref{equ:sca} and $\mathbf{a}(\eta_n^-)$ stands for $\mathbf{a}(\eta_n^-, \phi, \theta)$ derived by substituting $\eta_n^+$ with $\eta_n^-$. The resulting outputs of all $2N$ sub-arrays in $N$ sUCAs are expressed by $\mathbf{r}(\phi,\theta)\in\mathbb{C}^{2N\times1}$ which is given by:
\begin{equation}
\begin{aligned}
		\mathbf{r}(\phi,\theta)&=\mathbf{A}^T(\phi,\theta)\boldsymbol{\delta}\\
		&=\left[\text{AF}(\eta_0^+,\phi,\theta),\cdots,\text{AF}(\eta_{N-1}^+,\phi,\theta),\right.\\
		&\quad\enspace\left.\text{AF}(\eta_0^-,\phi,\theta),\cdots,\text{AF}(\eta_{N-1}^-,\phi,\theta)\right]^T,
\end{aligned}
	\label{equ:result_output}
\end{equation}
where $\text{AF}(\eta_n^+,\phi,\theta)$ and $\text{AF}(\eta_n^-,\phi,\theta)$ is the array factor of $\text{sub-array}_n^+$ or $\text{sub-array}_n^-$ given in \eqref{AF} and $\boldsymbol{\delta}$ is the phase shift vector in \eqref{phasev} introduced by variable delay line and is irrelevant to physical orientation. 

% \begin{figure}[!t]
%     \centering
%     \includegraphics[width=\linewidth]{cylinder_DCAA.pdf}
%     \caption{The overall scheme of communication system with cylinder DCAA. $2N$ sub-arrays are selectively connected to $N_\text{RF}$ RF chains through the selection matrix. }
%     \label{fig:cylinder_DCAA}
% \end{figure}
% The overall communication system with cylinder DCAA is shown in Fig. \ref{fig:cylinder_DCAA}. 
% To adaptively connect $2N$ sub-arrays to $N_{\text{RF}}$ RF chains, a \emph{selection matrix} denoted by $\mathbf{S}\in\{0,1\}^{N_{\text{RF}}\times2N}$ which satisfies $\Vert[\mathbf{S}]_{i,:}\Vert=1$ and $\Vert[\mathbf{S}]_{:,n}\Vert\leq1$, $1\leq i \leq N_{\text{RF}}$ and $1\leq n\leq 2N$ is proposed, where $N_{\text{RF}}$ is the number of RF chains. 
From the definition of \emph{selection matrix} given in Section \ref{sec:sys_model}, for a given RF chain tagged $i\in[1,N_\text{RF}]$, there always exists a sub-array tagged $n\in[1,2N]$, such that $[\mathbf{S}]_{i,n}=1$ and other elements in the row $i$ being 0. 
We can further define a set $\Omega$ that illustrates $N_\text{RF}$ chosen sub-arrays, which is given by:
\begin{equation}
	\Omega=\{n:\vert\vert[\mathbf{S}]_{:,n}\vert\vert=1\},
	\label{equ:omega_set}
\end{equation}
where $i,n\in\mathbb{Z}_+$. 
% The greedy optimization algorithm for the set $\Omega$ is given in \textbf{Algorithm} \ref{argo:find_S_f}. 
In the following two sections, we will present SINR performance of the uplink and downlink communication system with the proposed cylinder DCAA. 
\subsubsection{Uplink scenario}
For the uplink scenario of the proposed cylinder DCAA, the effective channel vector of the $k\text{th}$ user $\mathbf{h}_{\text{UP},k}\in\mathbb{C}^{2N\times1}$ is given by:
\begin{equation}
	\mathbf{h}_{\text{UP},k}=\sum_{l=1}^{L_k}\alpha_{k,l}\mathbf{r}(\phi_{k,l},\theta_{k,l}),
	\label{equ:effective_channel}
\end{equation}
where $L_k$ is the number of multipath components (MPCs) for the $k\text{th}$ user and $\alpha_{k,l}$ is 
the complex attenuation parameter of the $l\text{th}$ MPC for the $k\text{th}$ user. 

The additive Gaussian noise of each antenna element is given by matrix $\mathbf{Z}=[\mathbf{z}_1,\mathbf{z}_2,\cdots,\mathbf{z}_{2N}]^T\in\mathbb{C}^{2N\times M}$, and $\mathbf{z}_n\sim \mathcal{CN}(0,\sigma^2\textbf{I}_M)$, $n=1,2,\cdots,2N$, where $\sigma^2$ is the noise power. 
For notational convenience, we can derive an equivalent additive noise vector $\mathbf{z}'\in\mathbb{C}^{2N\times1}$ for $2N$ sub-arrays which is given by:
\begin{equation}
	\mathbf{z}'=\mathbf{Z}\boldsymbol{\delta}.
	\label{equ:sum_noise}
\end{equation}
Thereby $\mathbf{z}'\sim \mathcal{CN}(0,M\sigma^2\textbf{I}_{2N})$ is the equivalent sum noise vector over the $M$ elements of the $2N$ sub-arrays. 
The resulting uplink signal for the $k\text{th}$ user at the cylinder DCAA BS side is given by:
\begin{equation}
	y_{\text{UP},k}=\underbrace{\mathbf{w}_k^H\mathbf{S}\mathbf{h}_{\text{UP},k}s_k}_{k\text{th}\text{ user's signal}}+\underbrace{\mathbf{w}_k^H\mathbf{S}\sum_{\substack{i\neq k}}\mathbf{h}_{\text{UP},i}s_i}_{\text{inter-user interference}}+\underbrace{\mathbf{w}_k^H\mathbf{S}\mathbf{z}'}_{\text{additive noise}},
	\label{equ:result_signal_k}
\end{equation}
where $\mathbf{w}_k\in\mathbb{C}^{N_\text{RF}\times1}$ is the baseband beamforming vector of the $k\text{th}$ user, $\mathbf{z}'$ is the equivalent sum noise vector given in \eqref{equ:sum_noise}, $s_k$ is the information-bearing symbol of the $k\text{th}$ user that satisfies $\mathbb{E}[\vert s_k\vert^2]=P_{\text{UP},k}$, where $P_{\text{UP},k}$ is the transmit power of the $k\text{th}$ user. Hence, $\text{SINR}$ of the $k\text{th}$ user of cylinder DCAA is given by:
\begin{equation}
	\text{SINR}_{\text{UP},k}=\frac{
	\bar{P}_{\text{UP},k}\vert\mathbf{w}_k^H\mathbf{S}\mathbf{h}_{\text{UP},k}\vert^2
	}{
	\sum_{i\neq k}\bar{P}_{\text{UP},i}\vert\mathbf{w}_k^H\mathbf{S}\mathbf{h}_{\text{UP},i}\vert^2+M\Vert\mathbf{w}_k^H\mathbf{S}\Vert^2
	},
	\label{equ:SINR_cylinder}
\end{equation}
where $\bar{P}_{\text{UP},k}=P_{\text{UP},k}/\sigma^2$ is the transmit SNR of the $k$th user. 
The achievable maximum sum rate of $K$ users is given by:
\begin{equation}
	R_{\text{sum},\text{UP}}(\Omega)=\sum_{k=1}^{K}\log_2\left(
	1+\text{SINR}_{\text{UP},k}
	\right).
	\label{equ:Rsum}
\end{equation}
We aim to maximize the sum rate $R_{\text{sum},\text{UP}}(\Omega)$ of the cylinder DCAA for $K$ users, and the optimization problem is given by:
\begin{equation}
	\begin{aligned}
		\underset{\mathbf{S},\mathbf{W}}{\max} & \quad
		R_{\text{sum},\text{UP}}(\Omega)\\
		\text{s.t.} & \quad (\text{C1}):
		\enspace\mathbf{S}\in\{0,1\}^{N_{\text{RF}}\times2N}\\
		& \quad (\text{C2}):
		\enspace\Vert[\mathbf{S}]_{i,:}\Vert=1,\quad1\leq i\leq N_{\text{RF}}\\
		& \quad (\text{C3}):
		\enspace\Vert[\mathbf{S}]_{:,n}\Vert\leq1,\quad1\leq n\leq2N\\
		& \quad (\text{C4}):
		\enspace\Omega\enspace\text{ given in (25)}\\
		& \quad (\text{C5}):
		\enspace \mathbf{W}=\left[
		\mathbf{w}_1,\mathbf{w}_2,\cdots,\mathbf{w}_K
		\right]
	\end{aligned}
	\label{equ:optimization}
\end{equation}
where $\mathbf{W}\in\mathbb{C}^{N_{\text{RF}}\times K}$ is the baseband beamforming matrix consisting of $K$ baseband beamforming vectors for all users. 
% The greedy algorithm to find $\mathbf{S}$ and $\mathbf{W}$ in the uplink communication scenario is given by \cite{RAA_long} in \textbf{Algorithm} \ref{argo:find_S_f}. 
% \begin{algorithm}[htbp]
% 	\caption{Greedy algorithm for cylinder DCAA's selection matrix $\mathbf{S}$ and baseband beamforming matrix $\mathbf{W}$ in the uplink communication scenario.}
% 	\label{argo:find_S_f}
% 	\begin{algorithmic}[1]
% 		\STATE {Initialize $\Omega^0=\emptyset$ and $n\in\mathcal{U}^0=\{1,2,\cdots,2N\}$}
% 		\FOR{$i=1:N_{\text{RF}}$}
% 		\STATE {Calculate $n^i=\underset{n\in\mathcal{U}^{i-1}}{\text{argmax}}\quad R_{\text{sum},\text{UP}}\left(\Omega^{i-1}\bigcup n\right)$; }
% 		\STATE  {Update $\Omega^i=\Omega^{i-1}\bigcup n^i$;}
% 		\STATE {Update $\mathcal{U}^{i}=\mathcal{U}^{i-1}\setminus n^i$;}
% 		\ENDFOR
% 		\STATE {Obtain selection matrix $\mathbf{S}$ based on $\Omega^{N_{\text{RF}}}$;}
% 		\STATE {Obtain baseband beamforming matrix $\mathbf{W}$ based on $\mathbf{S}$ and equalization algorithm.}
% 		\STATE {\textbf{Output:} $\mathbf{S},\mathbf{W}, \text{and}$ $R_{\text{sum},\text{UP}}$}
% 	\end{algorithmic}
% \end{algorithm}
% What is worthy to be mentioned is that the typical underspread fading channel has the coherence time far larger than the delay spread \cite{wireless_comm}, where coherence time can be larger than $50 \mu\text{s}$ \cite{ref:coherence} which is sufficient for the greedy algorithm whose calculation complexity is $N_\text{RF}\log_2(N_\text{RF})$. 
For any given \emph{selection matrix} $\mathbf{S}$, the optimal receive beamforming is the linear minimum mean square error (MMSE) beamforming given by:
\begin{equation}
	\mathbf{w}_k(\mathbf{S})=\frac{\mathbf{C}_{\text{UP},k}^{-1}(\mathbf{S})\mathbf{S}\mathbf{h}_{\text{UP},k}}
    {
    \vert\vert\mathbf{C}_{\text{UP},k}^{-1}(\mathbf{S})\mathbf{S}\mathbf{h}_{\text{UP},k}\vert\vert
    },
	\label{equ:MMSE}
\end{equation}
where $\mathbf{C}_{\text{UP},k}(\mathbf{S})\in\mathbb{C}^{N_{\text{RF}}\times N_{\text{RF}}}$ is given by:
\begin{equation}
	\mathbf{C}_{\text{UP},k}(\mathbf{S})=\mathbf{S}\left(
	\sum_{i\neq k}P_{\text{UP},i}\mathbf{h}_{\text{UP},i}\mathbf{h}_{\text{UP},i}^H+M\sigma^2\mathbf{I}_{2N}
	\right)\mathbf{S}^T.
	\label{equ:C_matrix}
\end{equation}
Thus, \eqref{equ:Rsum} is reduced to:
\begin{equation}
    R_{\text{sum,UP}}(\Omega)=\sum_{k=1}^K\log_2(1+P_{\text{UP},k}(\mathbf{S}\mathbf{h}_k)^H\mathbf{C}_k^{-1}(\mathbf{S})(\mathbf{S}\mathbf{h}_k)).
    \label{equ:Rsum_reduced}
\end{equation}
The greedy algorithm to find $\mathbf{S}$ and $\mathbf{W}$ in the uplink communication scenario is given by \cite{RAA_long} in \textbf{Algorithm} \ref{argo:find_S_f}. 
% What is worthy to be mentioned is that the typical underspread fading channel has the coherence time far larger than the delay spread \cite{wireless_comm}, where coherence time can be larger than $50 \mu\text{s}$ \cite{ref:coherence} which is sufficient for the greedy algorithm whose calculation complexity is $N_\text{RF}\log_2(N_\text{RF})$, such that the algorithm can be performed within underspread channel's coherence time. 
\begin{algorithm}[htbp]
	\caption{Greedy algorithm to obtain cylinder DCAA's selection matrix $\mathbf{S}$ and baseband beamforming matrix $\mathbf{W}$ in the uplink scenario.}
	\label{argo:find_S_f}
	\begin{algorithmic}[1]
		\STATE {Initialize $\Omega^0=\emptyset$ and $n\in\mathcal{U}^0=\{1,2,\cdots,2N\}$}
		\FOR{$i=1:N_{\text{RF}}$}
		\STATE {Calculate $n^i=\underset{n\in\mathcal{U}^{i-1}}{\text{argmax}}\quad R_{\text{sum},\text{UP}}\left(\Omega^{i-1}\bigcup n\right)$; }
		\STATE  {Update $\Omega^i=\Omega^{i-1}\bigcup n^i$;}
		\STATE {Update $\mathcal{U}^{i}=\mathcal{U}^{i-1}\setminus n^i$;}
		\ENDFOR
		\STATE {Obtain selection matrix $\mathbf{S}$ based on $\Omega^{N_{\text{RF}}}$;}
		\STATE {Obtain baseband beamforming matrix $\mathbf{W}$ based on $\mathbf{S}$.}
		\STATE {\textbf{Output:} $\mathbf{S},\mathbf{W}, \text{and}$ $R_{\text{sum},\text{UP}}$}
	\end{algorithmic}
\end{algorithm}

\subsubsection{Downlink scenario}

For the downlink communication scenario with cylinder DCAA, the RF combiner of each sub-array needs to be replaced by a power splitter.
The downlink transmit information-bearing symbol vector for $K$ users is denoted by $\mathbf{s}_\text{DL}=\left[s_{\text{DL},1},\cdots,s_{\text{DL},K}\right]^T\in\mathbb{C}^{K\times1}$, which satisfies $\mathbb{E}[\mathbf{s}_\text{DL}\mathbf{s}^H_\text{DL}]=\mathbf{diag}([P_{\text{DL},1},\cdots,P_{\text{DL},K}])$, where $P_{\text{DL},k}$ denotes the transmit power for the $k\text{th}$ user. 
The downlink baseband precoding matrix for $K$ users is given by $\mathbf{W}_\text{DL}=\left[\mathbf{w}_{\text{DL},1},\mathbf{w}_{\text{DL},2},\cdots,\mathbf{w}_{\text{DL},K}\right]\in\mathbb{C}^{N_\text{RF}\times K}$ with $\mathbf{w}_{\text{DL},k}\in\mathbb{C}^{N_\text{RF}\times1}$ being the baseband precoding vector for the $k\text{th}$ user. 
Thus, after going through the baseband processing and the selection matrix $\mathbf{S}\in\mathbb{C}^{N_\text{RF}\times2N}$, the transmitted signal vector $\mathbf{x}_\text{DL}\in\mathbb{C}^{2N\times1}$ is given by:
\begin{equation}
\begin{aligned}
    	\mathbf{x}_\text{DL}&=\sqrt{\frac{1}{M}}\mathbf{S}^T\mathbf{W}_\text{DL}\mathbf{s}_\text{DL}
        =\sqrt{\frac{1}{M}}\mathbf{S}^T\sum_{k=1}^{K}\mathbf{w}_{\text{DL},k}s_{\text{DL},k},
\end{aligned}
	\label{equ:dl_x}
\end{equation}
and the factor $\sqrt{1/M}$ is introduced by the power splitter for the $M$ antenna elements in a sub-array. 
The downlink effective channel matrix for $K$ users is denoted by $\mathbf{H}_\text{DL}=\left[\mathbf{h}_{\text{DL},1},\mathbf{h}_{\text{DL},2},\cdots,\mathbf{h}_{\text{DL},K}\right]^T\in\mathbb{C}^{K\times2N}$, where $\mathbf{h}_{\text{DL},k}$ is the downlink effective channel for the $k\text{th}$ user: 
\begin{equation}
	\mathbf{h}_{\text{DL},k}=\sum_{l=1}^{L_k}\alpha_{k,l}\mathbf{r}\left(\phi_{k,l},\theta_{k,l}\right).
	\label{equ:dl_eff_channel}
\end{equation}
Thus, after the transmitted signal vector given in \eqref{equ:dl_x} passing through the effective channel, the received signal vector $\mathbf{y}_{\text{DL}}=\left[y_{\text{DL},1},y_{\text{DL},2},\cdots,y_{\text{DL},K}\right]^T\in\mathbb{C}^{K\times1}$ of $K$ users is given by:
\begin{equation}
	\begin{aligned}
	    \mathbf{y}_{\text{DL}}&=\mathbf{H}_\text{DL}\mathbf{x}_\text{DL}+\mathbf{z}_\text{DL}\\&=\sqrt{\frac{1}{{M}}}\mathbf{H}_\text{DL}\mathbf{S}^T\sum_{k=1}^{K}\mathbf{w}_{\text{DL},k}s_{\text{DL},k}+\mathbf{z}_\text{DL},
	\end{aligned}
	\label{equ:dl_y_vec}
\end{equation}
where $\mathbf{z}_\text{DL}=\left[z_1,z_2,\cdots,z_K\right]^T\sim\mathcal{CN}(0,\sigma^2\mathbf{I}_K)$ is the noise vector of $K$ users. The signal received by the $k\text{th}$ user is given by:
\begin{equation}
\begin{aligned}
	y_{\text{DL},k}&=\mathbf{h}_{\text{DL},k}^T\mathbf{x}_\text{DL}+z_k\\
	&=\sqrt{\frac{1}{M}}\mathbf{h}_{\text{DL},k}^T
	\left(
	\mathbf{S}^T\sum_{i=1}^{K}\mathbf{w}_{\text{DL},i}s_{\text{DL},i}
	\right)+z_k\\
	&=\underbrace{\sqrt{\frac{1}{M}}\mathbf{h}_{\text{DL},k}^T\mathbf{S}^T\mathbf{w}_{\text{DL},k}s_{\text{DL},k}}_{k\text{th user's signal}}+\\
	&\quad\underbrace{\sqrt{\frac{1}{M}}\mathbf{h}_{\text{DL},k}^T\mathbf{S}^T\left(\sum_{i\neq k}\mathbf{w}_{\text{DL},i}s_{\text{DL},i}\right)}_{\text{inter-user interference}}+z_k%\underbrace{z_k}_{\text{additive noise}}.
	\label{equ:dl_y}
\end{aligned}
\end{equation}
Based on \eqref{equ:dl_y}, the SINR of the cylinder DCAA at downlink scenario is given by:
\begin{equation}
	\text{SINR}_{\text{DL},k}=\frac{
	\frac{1}{M}P_{\text{DL},k}\vert\mathbf{h}_{\text{DL},k}^T\mathbf{S}^T\mathbf{w}_{\text{DL},k}\vert^2
	}{
	\frac{1}{M}\sum_{i\neq k}P_{\text{DL},i}\vert\mathbf{h}_{\text{DL},k}^T\mathbf{S}^T\mathbf{w}_{\text{DL},i}\vert^2+\sigma^2
	}.
	\label{equ:dl_sinr}
\end{equation}
Thus, the achievable maximum sum rate for downlink scenario of cylinder DCAA is given by:
\begin{equation}
	R_{\text{sum},\text{DL}}=\sum_{k=1}^{K}\log_2(1+\text{SINR}_{\text{DL},k}).
	\label{equ:dl_sum_rate}
\end{equation}
Specifically, from the uplink-downlink duality, for a given selection matrix $\mathbf{S}$, the optimal baseband transmit precoding is the MMSE precoding obtained from the dual uplink channels $\mathbf{H}_{\text{d-UP}}=\mathbf{H}_\text{DL}^H=[\mathbf{h}^*_{\text{DL},1},\mathbf{h}^*_{\text{DL},2},\cdots,\mathbf{h}^*_{\text{DL},K}]\in\mathbb{C}^{2N\times K}$ \cite{ref:duality_for_MIMO}. 
And with appropriate power allocation strategy, the same SINR and achievable maximum sum rate can be met at the downlink channel and its dual uplink channel with the same summarized transmit power from BS and user equipments (UEs), respectively. 
$\mathbf{w}_{\text{DL},k}$ for the $k\text{th}$ user in \eqref{equ:dl_y} is given by:
\begin{equation}
\mathbf{w}_{\text{DL},k}(\mathbf{S},\mathbf{p})=\frac{\mathbf{C}_{\text{DL},k}^{-1}(\mathbf{S},\mathbf{p})\mathbf{S}\mathbf{h}_{\text{DL},k}^*}{\vert\vert\mathbf{C}_{\text{DL},k}^{-1}(\mathbf{S},\mathbf{p})\mathbf{S}\mathbf{h}_{\text{DL},k}^*\vert\vert}%\frac{\mathbf{h}_{\text{DL},k}}{\sqrt{M}},
% \mathbf{w}_{\text{DL},k}=\sqrt{\bar{P}_\text{DL}}\frac{\mathbf{S}\mathbf{h}_{\text{DL},k}}{\vert\vert\mathbf{S}\mathbf{h}_{\text{DL},k}\vert\vert}, 
	\label{equ:dl_bf}
\end{equation}
where $\mathbf{p}=\left[{P_{\text{DL},1}},\cdots,{P_{\text{DL},K}}\right]^T\in\mathbb{R}^{K\times1}$ 
% is the power allocation vector given by:
% \begin{equation}
%     \mathbf{p}\vphantom{P_{\text{DL},1}}=\left[{P_{\text{DL},1}},\cdots,{P_{\text{DL},K}}\right]^T,
%     \label{equ:power_allo}
% \end{equation}
and $\mathbf{C}_{\text{DL},k}(\mathbf{S},\mathbf{p})\in\mathbb{C}^{N_\text{RF}\times N_\text{RF}}$ is given by:
\begin{equation}
    \mathbf{C}_{\text{DL},k}(\mathbf{S},\mathbf{p})=\mathbf{S}\left(
	\sum_{i\neq k}P_{\text{DL},i}%\frac{P_{\text{DL},i}}{M}
    \mathbf{h}_{\text{DL},i}^*\mathbf{h}_{\text{DL},i}^T+M\sigma^2\mathbf{I}_{2N}
	\right)\mathbf{S}^T.
\end{equation}
We aim to maximize the downlink achievable sum rate $R_{\text{sum,DL}}$ given in \eqref{equ:dl_sum_rate} by adjusting \emph{selection matrix} $\mathbf{S}$, baseband precoding matrix $\mathbf{W}_\text{DL}$ and power allocation vector $\mathbf{p}$ given in \eqref{equ:power_allocation}. 
Thus, the proposed optimization problem is given by:
\begin{equation}
    \begin{aligned}
		\underset{\mathbf{S},\mathbf{W}_\text{DL},\mathbf{p}}{\max} & \quad
		R_{\text{sum},\text{DL}}\\
		\text{s.t.} & \quad (\text{C1})-(\text{C3}) \text{ in (31)}\\
		& \quad (\text{D1}) \quad
		\mathbf{p}^T\mathbf{1}_{K\times1}=P_{\text{DL}},\\
    \end{aligned}
    \label{equ:dl_optimization}
\end{equation}
where $P_\text{DL}$ is the summarized transmit power in the downlink communication and $\frac{P_\text{DL}}{K\sigma^2}$ is defined as the average transmit SNR. 
However, owing to the binary constraints in \emph{selection matrix} $\mathbf{S}$ and the tight coupling between $\mathbf{S}$, $\mathbf{W}_\text{DL}$ and $\mathbf{p}$, problem \eqref{equ:dl_optimization} is difficult to be solved. 
Therefore, an iterative step-by-step optimization method for $\mathbf{S}$, $\mathbf{W}_\text{DL}$ and $\mathbf{p}$ is developed. 
For each iteration, $\mathbf{S}$ is updated firstly using the greedy algorithm described in \textbf{Algorithm} \ref{argo:find_S_f} with $\mathbf{H}_{\text{d-UP}}$. 
Then, $\mathbf{p}$ and $\mathbf{W}_\text{DL}$ w.r.t. $\mathbf{S}$ are optimized using the classic iterative waterfilling method \cite{ref:waterfilling}. 
This process is repeated until convergence or meeting the maximum times of iterations.
With waterfilling power allocation, the $k$th entry of $\mathbf{p}$ is given by: 
% To achieve better performance, a waterfilling power allocation method is introducted to find the optimal $\mathbf{p}$ with given \emph{selection matrix} $\mathbf{S}$ and baseband beamforming matrix $\mathbf{W}_\text{DL}$ and the $k\text{th}$ entry of $\mathbf{p}$ is given by: 
\begin{equation}
    \begin{aligned}
    P_{\text{DL},k}\approx\max\left(\mu-\frac{Z_k}{\vert\mathbf{h}^T_{\text{DL},k}\mathbf{S}^T\mathbf{w}_{\text{DL},k}\vert^2},0\right),
    \end{aligned}
    \label{equ:power_allocation}
\end{equation}
where $Z_k$ is equivalent interference-plus-noise power which is given by:
% where $\mathbf{\Gamma}\in\mathbb{R}^{K\times K}$ has component $\mathbf{\Gamma}_{k,j}$ equals to:
\begin{equation}
Z_k=\sum_{j\neq k}P_{\text{DL},j}\vert\mathbf{h}^T_{\text{DL},k}\mathbf{S}^T\mathbf{w}_{\text{DL},j}\vert^2+M\sigma^2,
% \mathbf{\Gamma}_{k,j}=
% \begin{cases}
%     & \frac{\vert\mathbf{h}^T_{\text{DL},k}\mathbf{S}^T\mathbf{{w}}_{\text{DL},k}\vert^2}{\text{SINR}_{\text{DL},k}M\sigma^2}\quad(k=j)\\
%     & -\frac{\vert\mathbf{h}^T_{\text{DL},k}\mathbf{S}^T\mathbf{{w}}_{\text{DL},j}\vert^2}{M\sigma^2}.\quad(k\neq j).
% \end{cases}
    % a_k=\frac{(1+\text{SINR}_{\text{DL},k})\vert\mathbf{h}^T_{\text{DL},k}\mathbf{S}^T\mathbf{w}_{\text{DL},k}\vert^2}{
    % \text{SINR}_{\text{DL},k}
    % },
    \label{equ:a_k_in_p}
\end{equation}
and $\mu$ is the water level given by:
\begin{equation}
    \mu=\frac{P_\text{DL}}{K}+\frac{1}{K}\sum_{k=1}^{K}\frac{Z_k}{\vert\mathbf{h}^T_{\text{DL},k}\mathbf{S}^T\mathbf{w}_{\text{DL},k}\vert^2}.
    \label{equ:power_constraint}
\end{equation}

While iteratively updating $\mathbf{S}$ based on the updated $\mathbf{p}$ with \textbf{Algorithm} \ref{argo:find_S_f} ensures accuracy, its computational complexity is over $\mathcal{O}(NN_\text{RF}\log(\frac{1}{\varepsilon_\text{th}}))$, where $\varepsilon_\text{th}$ is the convergence threshold.
The complexity is intolerable for instant communication considering channel coherency. 
To address this challenge, we propose a simplification where $\mathbf{S}$ is updated only once during the initialization phase with uniform power allocation, i.e., $P_{\text{DL},k}=\frac{P_\text{DL}}{K}, k=1,2,\cdots,K$. 
This trade-off between computational efficiency and accuracy is analyzed in Section \ref{sub_sec:eff_and_conv}, where it is demonstrated that the proposed method achieves a substantial reduction in computational complexity with only a little compromising of performance. 
% However, \textbf{Algorithm} \ref{argo:find_S_f} is repeatedly executed for every iteration to determine a suitable $\mathbf{S}$ based on the updated $\mathbf{p}$, which results in bulky computational complexity. 
% To substantially reduce the complexity with little performance compromise, $\mathbf{S}$ is updated only once with the initial uniform power allocation, i.e., $P_{\text{DL},k}=\frac{P_\text{DL}}{K}, k=1,2,\cdots,K$. 
% As a result, during the iterative process, only $\mathbf{p}$ and $\mathbf{W}_\text{DL}$ are updated while $\mathbf{S}$ remains fixed. 
% However, we need to perform \textbf{Algorithm} \ref{argo:find_S_f} per iteration to find a suitable $\mathbf{S}$ with an updated $\mathbf{p}$, which will introduce bulky computational complexity. 
% To sharply decrease complexity with little performance compromising, the selection matrix $\mathbf{S}$ is only updated with the initialized uniform power allocation, i.e., $P_{\text{DL},k}=\frac{P_\text{DL}}{K}, k=1,2,\cdots,K$. 
% Therefore, the iteration process will only update $\mathbf{p}$ and $\mathbf{W}_\text{DL}$ with fixed $\mathbf{S}$. 
Finally, the proposed downlink optimization algorithm is given by \textbf{Algorithm} \ref{algo:dl_optim}. 
\begin{algorithm}[htbp]
    \caption{Algorithm to obtain cylinder DCAA’s \emph{selection matrix} $\mathbf{S}$, baseband beamforming matrix $\mathbf{W}_\text{DL}$ and power allocation vector $\mathbf{p}$ in the downlink communication.}
    \label{algo:dl_optim}
    \begin{algorithmic}[1]
    \STATE{\textbf{Input:} Maximum iteration times $T_\text{max}$, convergence threshold $\varepsilon_\text{th}$, downlink effective channel $\mathbf{H}_\text{DL}$ and summarized power $P_\text{DL}$;}
    \STATE{\textbf{Initialize:} $\mathbf{p}^{(0)}=\frac{P_\text{DL}}{K}\mathbf{1}_{K\times1}$, $t=0$;}
    \STATE{Obtain $\mathbf{S},\mathbf{W}^{(0)}_\text{DL}$ from \textbf{Algorithm} \ref{argo:find_S_f};}
    \REPEAT
    \STATE{Update $t\leftarrow t+1$;}
    \STATE{Obtain $\mathbf{W}^{(t)}_\text{DL}$ from \eqref{equ:dl_bf};}
    % \FOR{$i=1,2,\cdots,K$}
    % \STATE{Calculate $\text{SINR}^{(t)}_{\text{DL},i}$ with $\mathbf{W}^{(t)}_\text{DL}$ from \eqref{equ:dl_sinr};}
    % \ENDFOR
    \STATE{Obtain $\mathbf{p}^{(t)}$ from \eqref{equ:power_allocation};}
    \IF{$t\geq T_\text{max}$}
    \STATE{break;}
    \ENDIF
    \UNTIL{Convergence: $(\vert\mathbf{p}^{(t)}-\mathbf{p}^{(t-1)}\vert)^T\mathbf{1}_{K\times1}<\varepsilon_\text{th}$}
    \STATE{Calculate $\text{SINR}_{\text{DL},i}, i=1,2,\cdots,K$  from \eqref{equ:dl_sinr};}
    \STATE{\textbf{Output:} $\mathbf{S},\mathbf{W}^{(t)}_\text{DL},\mathbf{p}^{(t)},$ and $R_{\text{sum,DL}}$;}
    \end{algorithmic}
\end{algorithm}

\subsection{The benchmark: ULA with HBF structure}

\begin{figure}[htbp]
	\centering
	\includegraphics[width=0.8\linewidth]{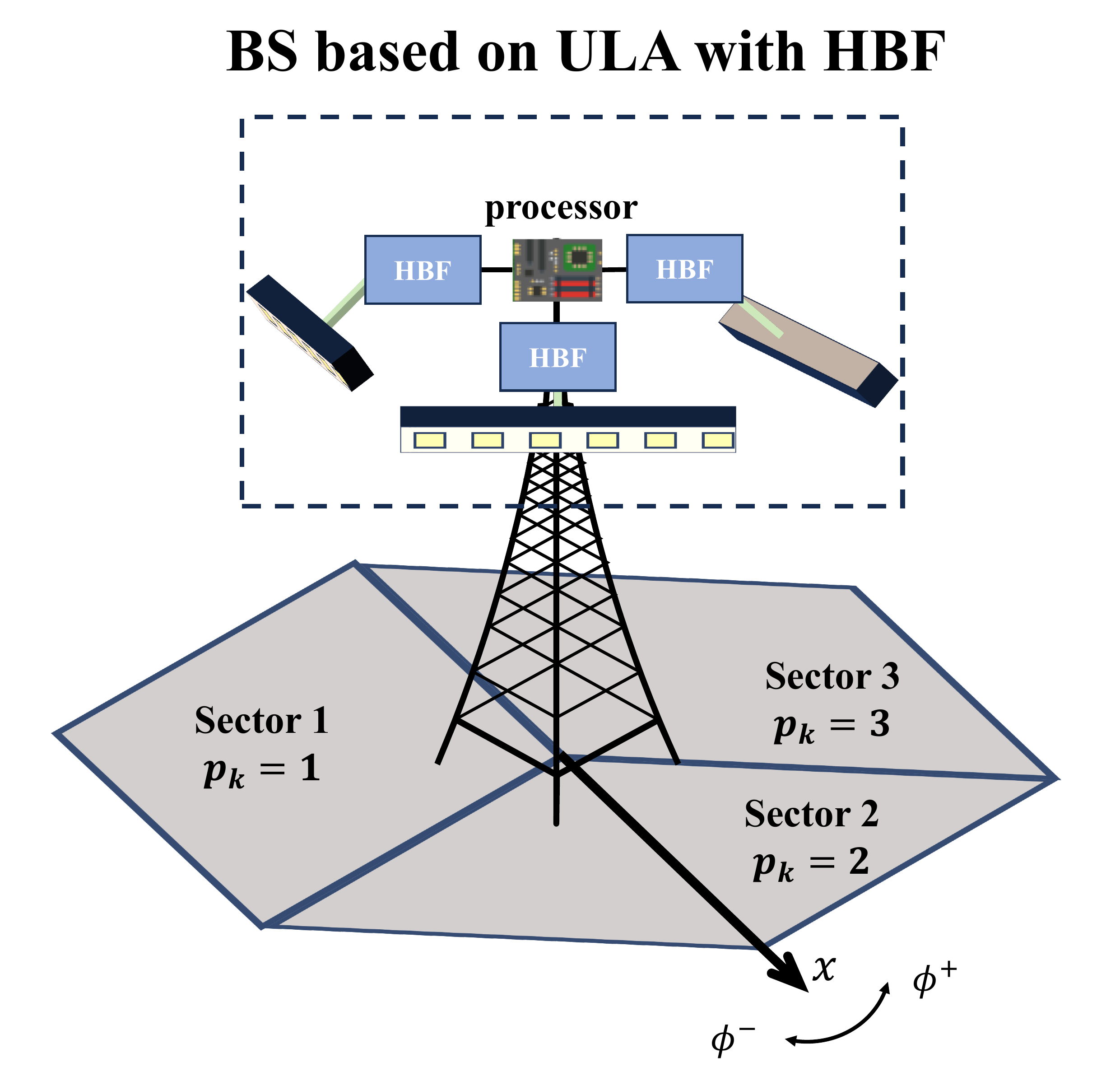}
	\caption{The cell sectoring communication system where BS is equipped with ULA and uses HBF to achieve beam steering.}
	\label{fig:HBF}
\end{figure}

As illustrated in Fig. \ref{fig:HBF}, we consider a widely implemented cell sectoring system based on ULA in 3GPP technical report \cite{3GPP} with classic HBF as the benchmark. 
The cell sectoring system based on ULA with HBF is operated by dividing the cell into 3 sectors, and each covers an angle interval of $120^\circ$. 
The BS aligns a ULA with DFT codebook to serve users in the specific sector. Let $p_k\in\{1,2,3\}$ refers to the sector number of the $k\text{th}$ user determined by the AoA $\phi_k$ which is defined as: 
\begin{equation}
	p_k=
	\begin{cases}
		&1\qquad(-\pi<\phi_k\leq-\frac{\pi}{3})\\
		&2\qquad(-\frac{\pi}{3}<\phi_k\leq\frac{\pi}{3})\\
		&3\qquad(\frac{\pi}{3}<\phi_k\leq\pi). 
	\end{cases}
	\label{equ:sector_def}
\end{equation} 
The response vector of the ULA that serves the sector $p_k$ is given by:
\begin{equation}
\begin{aligned}
		&\mathbf{a}_{\text{ULA}}(p_k,\phi)=\\
		&\sqrt{G\left(\xi(p_k),\frac{\pi}{2}\right)}\left[1,e^{-j\pi\sin\xi(p_k)},\cdots,e^{-j\pi(M-1)\sin\xi(p_k)}\right]^T,
	\label{equ:ULA_cell_sectoring}
\end{aligned}
\end{equation}
where the relative azimuth angle $\xi(p_k)=\phi-(p_k-2)\frac{2\pi}{3}$. The effective channel of the $k\text{th}$ user is given by:
\begin{equation}
	\mathbf{h}_{\text{UP}_\text{ULA},k}=\sum_{l=1}^{L_k}\alpha_{k,l}\mathbf{a}_\text{ULA}(p_k,\phi_{k,l}),
	\label{equ:cell_sectoring_response_vector}\end{equation}
where $L_k$ is the number of paths of the $k\text{th}$ user. 
The DFT codebook analog beamforming matrix $\mathbf{F}\in\mathbb{C}^{N_\text{RF}\times M}$ is given by: 
\begin{equation}
\begin{aligned}
	\mathbf{F}=\mathbf{\Xi}\mathbf{U},
	\label{equ:cell_sectoring_F}
\end{aligned}
\end{equation}
where $\mathbf{U}=[u_{l,m}]\in\mathbb{C}^{M\times M}$, $0\leq l,m\leq M-1$ is the DFT matrix with $u_{l,m}=\exp(-j2\pi\frac{lm}{M})/\sqrt{M}$ and $\mathbf{\Xi}=\left(\mathbf{n}_1,\mathbf{n}_2,\cdots,\mathbf{n}_{N_{\text{RF}}}\right)^T$ is the digital selection matrix and the vector $\mathbf{n}_i, 1\leq i\leq N_{\text{RF}}$ has only one element being 1 and zero elsewhere. 
Given the substantial angular separation between different sectors and the deployment of directional antenna elements, the inter-sector interference is considered negligible compared to the intra-sector interference. 
% Consequently, we focuses solely on the intra-sector inter-user interference. 
Thus, the received signal solely focused on the intra-sector inter-user interference is given by:
\begin{equation}
    \begin{aligned}
y_{\text{UP}_\text{ULA},k}=\mathbf{w}_{\text{ULA},k}^H\mathbf{F}\left(\sum_{\substack{i=1\\p_i=p_k}}^{K}\mathbf{h}_{\text{UP}_\text{ULA},i}s_{i}+\mathbf{z}\right), 
    \end{aligned}
    \label{equ:ULA_received}
\end{equation}
where $\mathbf{z}\sim\mathcal{CN}(0,\sigma^2\mathbf{I}_M)$ is the additive noise and $\mathbf{w}_{\text{ULA},k}$ is the digital MMSE baseband beamforming vector which is given by: 
\begin{equation}
    \begin{aligned}
    \mathbf{w}_{\text{ULA},k}=P_{\text{UP},k}\mathbf{C}^{-1}_{\text{ULA},k}(\mathbf{F})\mathbf{F}\mathbf{h}_{\text{UP}_\text{ULA},k}.
    \end{aligned}
    \label{equ:ULA_BF}
\end{equation}
where $\mathbf{C}_{\text{ULA},k}(\mathbf{F})\in\mathbb{C}^{N_\text{RF}\times N_\text{RF}}$ is given by:
\begin{equation}
\begin{aligned}
    \mathbf{C}_{\text{ULA},k}(\mathbf{F})&=\sum_{\substack{i\neq k\\p_i=p_k}}P_{\text{UP},i}\left(\mathbf{F}\mathbf{h}_{\text{UP}_\text{ULA},i}\right)\left(\mathbf{F}\mathbf{h}_{\text{UP}_\text{ULA},i}\right)^H+\mathbf{F}\mathbf{F}^H\sigma^2\\
    &=\mathbf{F}\left(
    \sum_{\substack{i\neq k\\p_i=p_k}}P_{\text{UP},i}\mathbf{h}_{\text{UP}_\text{ULA},i}\mathbf{h}^H_{\text{UP}_\text{ULA},i}+\sigma^2\mathbf{I}_{M}
    \right)\mathbf{F}^H.
\end{aligned}
\label{equ:ULA_MMSE}
\end{equation}
Optimization process of digital selection matrix $\mathbf{\Xi}$ is given in \cite{ref:HBF_DFT}. Thus, SINR of the $k\text{th}$ user in the cell sectoring system based on ULA with HBF is given by:
\begin{equation}
    \begin{aligned}
	&\text{SINR}_{\text{UP}_\text{ULA},k}\\
    &=\frac{
	P_{\text{UP},k}\left\vert\mathbf{w}^H_{\text{ULA},k}\mathbf{F}\mathbf{h}_{\text{UP}_\text{ULA},k}\right\vert^2
	}{
	\sum_{\substack{i\neq k\\p_i=p_k}}P_{\text{UP},i}\left\vert\mathbf{w}^H_{\text{ULA},k}\mathbf{F}\mathbf{h}_{\text{UP}_\text{ULA},i}\right\vert^2+\sigma^2\left\vert\left\vert\mathbf{F}^H\mathbf{w}_{\text{ULA},k}\right\vert\right\vert^2
	}.
    \end{aligned}
    \label{equ:DL_ULA_SINR}
\end{equation}
Therefore, the achievable maximum sum rate of the benchmark system in the uplink communication scenario is given by:
\begin{equation}
	R_{\text{sum},\text{UP}_\text{ULA}}=\sum_{k=1}^{K}\log_2\left(1+\text{SINR}_{\text{UP}_\text{ULA},k}\right).
	\label{equ:cell_sectoring_sum_rate}
\end{equation}

Similarly, for cell sectoring system based on ULA with HBF in the downlink scenario, the $k\text{th}$ user's received signal is given by:
\begin{equation}
	y_{\text{DL}_\text{ULA},k}=\sum_{\substack{i=1\\p_i=p_k}}^K\mathbf{h}^T_{\text{DL}_\text{ULA},k}\mathbf{F}^H\mathbf{w}_{\text{DL}_\text{ULA},i}s_{\text{DL},i}+z_k,
	\label{equ:dl_ula_y}
\end{equation}
where $\mathbf{h}_{\text{DL}_\text{ULA},k}\in\mathbb{C}^{M\times1}$ is the downlink effective channel for the $k\text{th}$ user at sector denoted by $p_k$, $\mathbf{w}_{\text{DL}_\text{ULA},i}\in\mathbb{C}^{N_\text{RF}\times1}$ is the transmit MMSE baseband beamforming vector of the $i\text{th}$ user obtained from uplink-downlink duality \cite{ref:duality_for_HBF} and the power splitting factor $\sqrt{1/M}$ is implied in the analog beamforming matrix $\mathbf{F}$. 
Thus, the SINR of the cell sectoring based on ULA with HBF is given by:
\begin{equation}
\begin{aligned}
	&\text{SINR}_{\text{DL}_\text{ULA}, k}\\
    &=\frac{
	P_{\text{DL},k}\left\vert\mathbf{h}^T_{\text{DL}_\text{ULA},k}\mathbf{F}^H\mathbf{w}_{\text{DL}_\text{ULA},k}\right\vert^2
	}{
	\sum_{\substack{i\neq k\\p_i=p_k}}P_{\text{DL},i}\left\vert\mathbf{h}^T_{\text{DL}_\text{ULA},k}\mathbf{F}^H\mathbf{w}_{\text{DL}_\text{ULA},i}\right\vert^2+\sigma^2
	},
    \end{aligned}
	\label{equ:dl_ula_sinr}
\end{equation}
and the achievable maximum downlink sum rate of cell sectoring based on ULA with HBF is given by:
\begin{equation}
	R_{\text{sum},\text{DL}_\text{ULA}}=\sum_{k=1}^{K}\log_2\left(1+\text{SINR}_{\text{DL}_\text{ULA},k}\right).
	\label{equ:dl_ula_sum_rate}
\end{equation} 

\section{Simulation and numerical results}

% The response pattern of cylinder DCAA shown in Fig. \ref{cylinder} with $M=16$ directional antenna elements and $N=13$ works under central frequency $f_c=47.2\text{GHz}$ is shown in Fig. \ref{BS_res}. 

% The numerical results in Fig. \ref{BS_res} illustrates that the cylinder DCAA 
% achieves uniform spatial resolution without any beamforming technology, where $\text{sub-arrays}^+$ uniformly cover the positive angle domain $\phi\in(0,\pi]$ and $\text{sub-arrays}^-$ uniformly cover the negative angle domain $\phi\in(-\pi,0]$ together with relatively low sidelobe level due to the directional antenna elements. The design of orientation angle $\eta_n^+$ and $\eta_n^-$ ensures that the sub-arrays will have as little interference as possible among vertically adjacent sub-arrays. 

% \begin{figure}[!t]
% 	\centering
% 	\includegraphics[width=\linewidth]{response2.pdf}
% 	\caption{Response pattern of cylinder DCAA, whose $M=16$, $N=13$ and the system works under central frequency $f_c=47.2\text{GHz}$. Cylinder DCAA achieves uniform spatial resolution with multiples sUCAs arranged with different physical orientations. }
% 	\label{BS_res}
% \end{figure}

\begin{figure}[!t]
	\centering
	\subfloat[Uplink communication]{%
		\includegraphics[width=\linewidth]{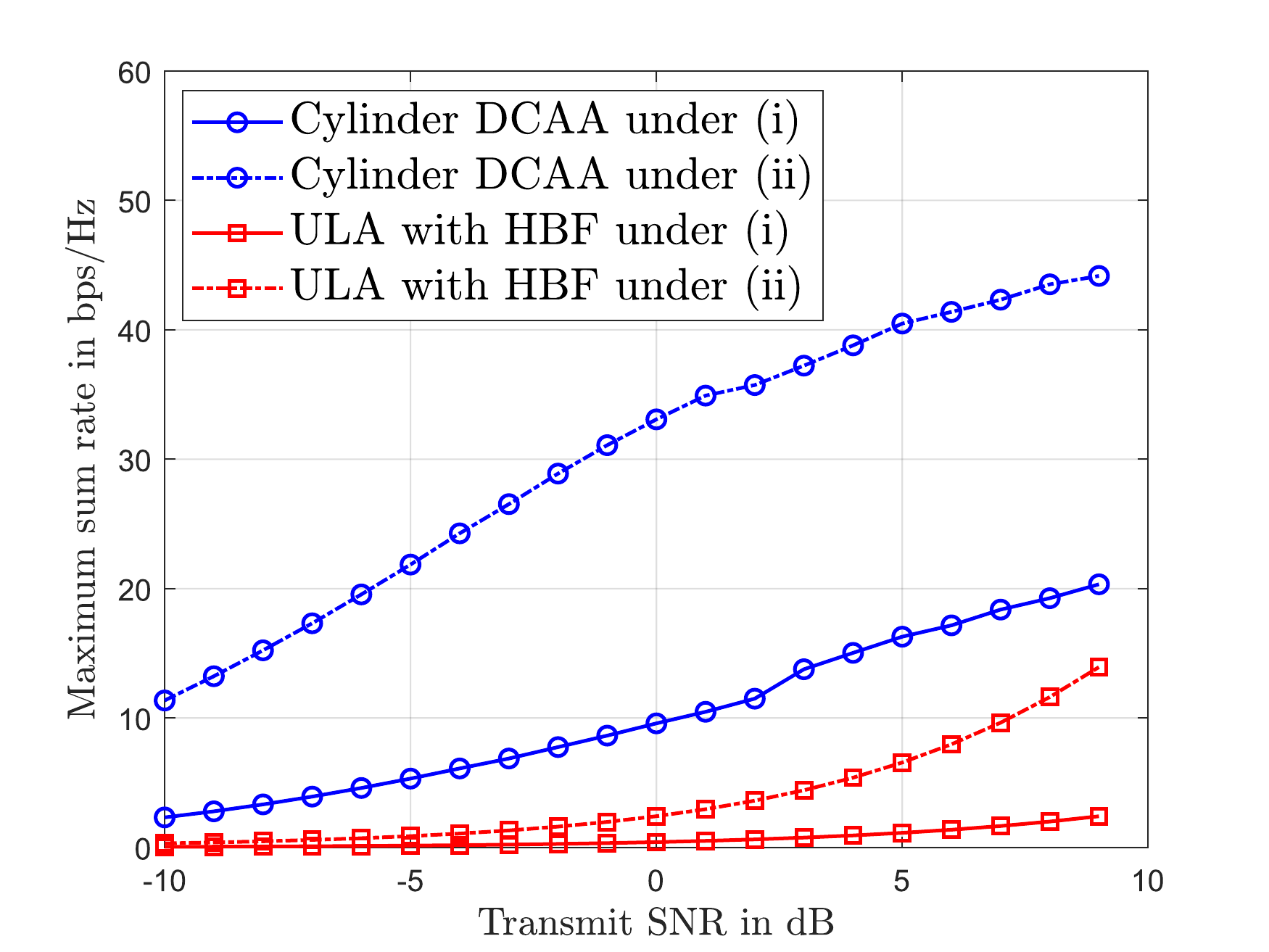}
		\label{fig:sum_rate1}
	}
	\hfil
	\subfloat[Downlink communication]{%
		\includegraphics[width=\linewidth]{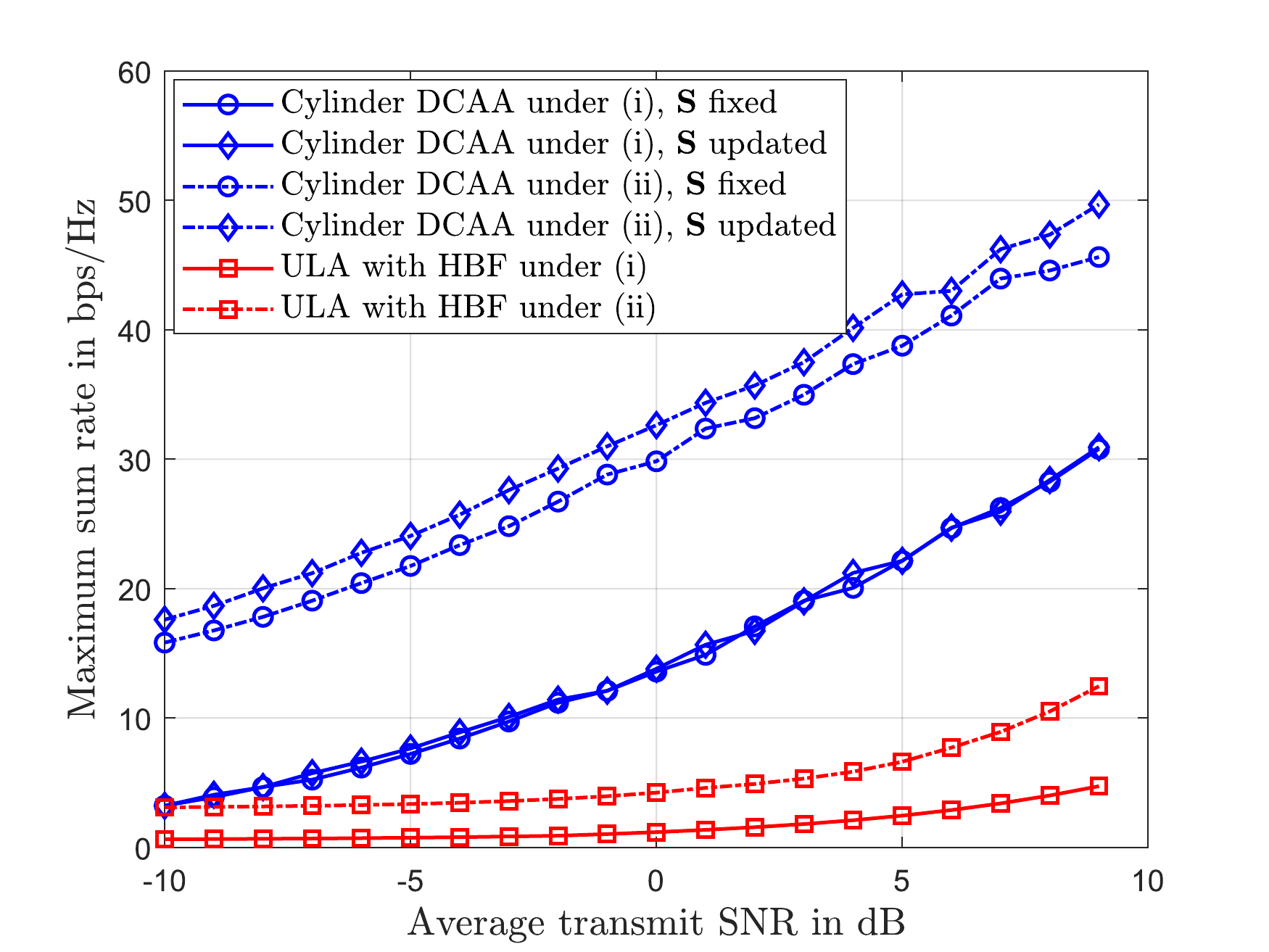}
		\label{fig:sum_rate2}
	}
	\caption{Uplink and downlink maximum sum rate performance comparison of the communication system with cylinder DCAA and the cell sectoring system based on ULA with HBF under 2 scenarios. 
    The communication system with cylinder DCAA achieves higher maximum sum rate performance with same transmit SNR, especially in scenario (ii).} 
	\label{fig:R_sum}
\end{figure}

\begin{figure}[!t]
    \centering
    \includegraphics[width=\linewidth]{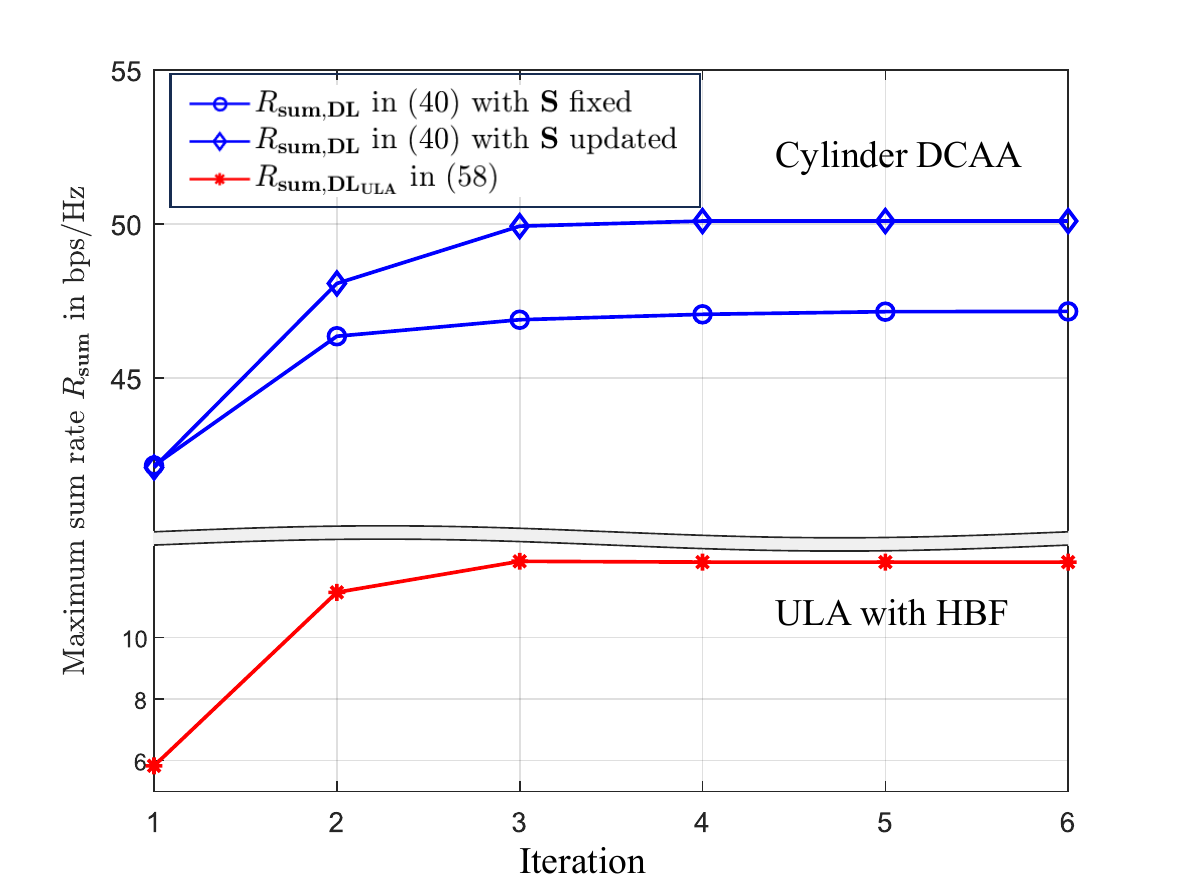}
    \caption{Convergence of \textbf{Algorithm} \ref{algo:dl_optim} for cylinder DCAA and ULA with HBF at the dense connectivity scenario, where $M=128$, $K=N_\text{RF}=30$ and $\varepsilon_\text{th}=0.1$ under average transmit SNR being $9$ dB. }
    \label{fig:conv}
\end{figure}

\subsection{Uplink and downlink sum rate performance of cylinder DCAA}\label{sub_sec:simu}
In this part, we present simulation results of the uplink and downlink maximum sum rate performance of the communication system with cylinder DCAA and compare it with the benchmark cell sectoring system based on ULA with HBF. 
Here we choose two typical communication scenarios: (i) normal connectivity with 10 users and (ii) dense connectivity with 30 users in a cell, detailed parameters are given in Table \ref{tab:scenario}.
\begin{table}[htbp]
	\centering
	\caption{Simulation scenarios. }
	\label{tab:scenario}
	\begin{tabular}{|c|c|c|c|}
		\hline
        Scenario & \makecell{$M$: antennas \\per sub-array or ULA } & \makecell{$N_{\text{RF}}$:\\ RF chains} & \makecell{$K$: users}\\
        \hline
        \makecell{(i) \\ normal connectivity} & 64 & 10 & 10\\ 
        \hline
        \makecell{(ii) \\ dense connectivity} & 128 & 30 & 30\\
        \hline
        \multicolumn{4}{|c|}{Antenna pattern: 3GPP directional antenna \cite{3GPP}}\\
        \hline
	\end{tabular}
\end{table}
% In the first scenario, communication system with cylinder DCAA has $M=64$ antenna elements in a sub-array and the benchmark cell sectoring system based on ULA with HBF has $M=64$ antenna elements in a ULA. 
% In the second scenario, communication system with cylinder DCAA has $M=128$ antenna elements in a sub-array and cell sectoring system based on ULA with HBF has $M=128$ antenna elements in a ULA. 
% Communication system with cylinder DCAA and cell sectoring system based on ULA with HBF in both scenarios have directional antenna elements that satisfies 3GPP 38.901 technical specification \cite{3GPP}. 
%As illustrated in Fig. \ref{fig:R_sum}, 
Wireless channel generation parameters for the simulation are shown in Table \ref{tab:simu}. 
% Simulation is based on the 3GPP in-door office scenario, which has the position of BS similar to that in proposition \ref{prop:farfield}. 
%In the benchmark scenario, $M=64$ and $M=128$ antenna elements are arranged into a ULA to serve a sector of $\frac{2\pi}{3}$ angle interval using DFT codebook. 
%The comparison of maximum sum rate performance of proposed cylinder architecture and the ULA-based cell sectoring system is illustrated in Fig. \ref{fig:R_sum}. 
As illustrated in Fig. \ref{fig:R_sum}(a), for the uplink scenario, communication system with cylinder DCAA demonstrates superior maximum sum rate performance compared to cell sectoring system based on ULA with HBF, and the performance advantage is evident under both scenario (i) and (ii), given the same transmit SNR at the user terminal. 
% Notably, the performance improvement becomes particularly pronounced in dense connectivity scenarios where the cylinder DCAA incorporates more antenna elements within each sub-array. 
% both in normal connectivity with $K=10$ users and dense connectivity with $K=30$ users with the same transmit SNR at user equipment (UE), especially in the dense connectivity scenario with more antenna elements in a sub-array. 
As illustrated in Fig. \ref{fig:R_sum}(b), 
% for the downlink communication, updating $\mathbf{S}$ provides only $8.2\%$ performance improvement comparing with fixed $\mathbf{S}$ method given in \textbf{Algorithm} \ref{algo:dl_optim} under scenario (ii). 
% Therefore, by updating $\mathbf{S}$ only once, we we can achieve significantly lower computational complexity while maintaining satisfactory performance. 
% This trade-off makes the single-update scheme a more practical choice in scenarios with limited computational resources. 
with the same average transmit SNR denoted by $\frac{P_{\text{DL}}}{K\sigma^2}$ at BS side, cylinder DCAA still achieves superior maximum sum rate in the downlink communication than the cell sectoring system based on ULA with HBF, thanks to its uniform spatial resolution and the enhanced beamforming gain facilitated by the cylinder DCAA architecture. 
% both communication system with cylinder DCAA and cell sectoring system based on ULA with HBF experience degraded performance in the low average transmit SNR region at the BS. 
% However, as the average transmit SNR increases, the cylinder DCAA system exhibits a significant enhancement in maximum sum rate performance. 
% This improvement can be attributed to 
% However, with the increasing of average transmit SNR, the maximal sum rate of communication system with cylinder DCAA has significant performance improvement, thanks to the uniform spatial resolution and enhanced beamforming gain. 
% But at the second dense connectivity scenario, where exists $K=30$ users, $N_\text{RF}=30$ RF chains and $M=128$ antenna elements in a sub-array or ULA, communication system with cylinder DCAA has significantly better performance than cell sectoring system based on ULA with HBF with the same transmit SNR in UE or BS, respectively, which is attributed to the uniform spatial resolution that suppress intra-sector inter-user interference, thus improving users SINR. 
The satisfying simulation results imply that the proposed cylinder DCAA is a potential cost-efficient solution for the high dense connectivity in the future 6G networks.

\subsection{Effectiveness and convergence of the proposed Algorithm 2}\label{sub_sec:eff_and_conv}

As illustrated in Fig. \ref{fig:R_sum}(b), updating $\mathbf{S}$ provides similar performance under scenario (i) and only $8.2\%$ performance improvement compared with the fixed $\mathbf{S}$ method given in \textbf{Algorithm} \ref{algo:dl_optim} under scenario (ii). 
Therefore, by updating $\mathbf{S}$ only once, we can achieve significantly lower computational complexity $\mathcal{O}(\log(\frac{1}{\varepsilon_\text{th}}))$ for the iteration process while maintaining satisfactory performance. 
This trade-off makes the single-update scheme a more practical choice for implementation. 

Fig. \ref{fig:conv} illustrates the convergence behavior of the proposed \textbf{Algorithm} \ref{algo:dl_optim} for the proposed cylinder DCAA in scenario (ii) under average transmit SNR being $9$ dB. 
Similarly, for cell sectoring system based on ULA with HBF, \textbf{Algorithm} \ref{algo:dl_optim} is adapted to adjust DFT-based HBF matrix $\mathbf{F}$, baseband beamforming matrix $\mathbf{W}_{\text{DL}_\text{ULA}}$ and power allocation vector $\mathbf{p}$ for larger maximum sum rate. 
We define the convergence threshold being $\varepsilon_\text{th}=0.01$ for the power allocation vector $\mathbf{p}$. 
As illustrated in Fig. \ref{fig:conv}, the proposed \textbf{Algorithm} \ref{algo:dl_optim} converges within $6$ iterations for both cylinder DCAA and ULA with HBF.   
These results jointly validate the effectiveness and convergence of the proposed optimization algorithm.

%The results are illustrated in Fig. \ref{fig:R_sum}.

%\begin{figure}[!t]
%	\centering
%	\includegraphics[width=\linewidth]{sum_rate.pdf}
%	\caption{Sum rate performance comparison of the proposed cylinder architecture and the ULA-based cell sectoring system, where $M=64$, $N=52$, $N_{\text{RF}}=K=10$ and $f_c=47.2\text{GHz}$. The proposed cylinder architecture achieves higher maximum sum rate performance with same transmit SNR together better cost-efficiency. }
%	\label{fig:R_sum}
%\end{figure}

\begin{table*}[htbp]
	\centering
	\caption{Simulation parameters in 3GPP Indoor-office NLoS scenario \cite{3GPP}.}
	\label{tab:simu}
	\begin{tabular}{|c|c|c|}
		\hline
		\textbf{Parameter} & \textbf{Symbol} & \textbf{Definition / Value} \\
		\hline
		Number of users & $K$ & 10 and 30\\
		\hline
		Number of RF chains & $N_{\text{RF}}$ & 10 and 30\\
		\hline
		LoS azimuth angle of the $k\text{th}$ user & $\phi_k$ & $\phi_k\sim \mathcal{U}(-\pi,\pi)$\quad($k=1,2,\cdots,K$)\\
		\hline
		LoS zenith angle of the $k\text{th}$ user & $\theta_k$ & $\theta_k=\pi/2$ \quad ($k=1,2,\cdots,K$)\\
        \hline
		Carrier central frequency & $f_c$ & 47.2\text{GHz}\\
		% \hline
		% Number of clusters per angle set & $N_c$ & 19\\
		% \hline
		% Number of rays per cluster & $N$ & 20\\
		\hline
		Number of paths of the $k\text{th}$ user & $L_k$ & $L_k=N_c\times N_r,\enspace N_c=19,N_r=20$\\
		\hline
		Cluster delay & $\tau_{n_c}$ & \makecell{
		$\tau_{n_c}=\text{sort}(\tau'_{n_c}-\text{min}(\tau'_{n_c}))$, $\tau'_{n_c}=-r_{\tau}\text{DS}\log(X_n)$\quad ($n_c=1,2,\cdots,N_c$)\\
		$\text{lgDS}\sim \mathcal{N}(-7.173-0.28\log_{10}(1+f_c),(0.10\log_{10}(1+f_c)+0.055)^2)$,\\ lgDS=$\log_{10}(\text{DS}/1\text{s})$, 
		$X_n\sim \mathcal{U}(0,1)$, $r_{\tau}=3$
		}\\
		\hline
		Cluster power & $P_{n_c}$ & \makecell{
		$P_{n_c}=P'_{n_c}/\sum_{n_c=1}^{N_c}P'_{n_c}$, $P'_{n_c}=\text{exp}\left(-\tau_{n_c}(r_{\tau}-1)/(r_{\tau}\text{DS})\right)10^{-Z_n/10}$\\
		$Z_n\sim \mathcal{N}(0,3^2)$, ($n_c=1,2,\cdots,N_c$)
		}\\
		\hline
		Azimuth angle spread & ASA & \makecell{
		$\text{lgASA}\sim \mathcal{N}(1.863-0.11\text{log}_{10}(1+f_c),(0.12\log_{10}(1+f_c)+0.059)^2)$,\\
		lgASA=$\text{log}_{10}(\text{ASA}/1^{\circ})$
		}\\
		\hline
		zenith angle spread & EAS & \makecell{
		$\text{lgEAS}\sim \mathcal{N}(1.387-0.15\text{log}_{10}(1+f_c),(-0.09\log_{10}(1+f_c)+0.746)^2)$,\\
		lgEAS=$\text{log}_{10}(\text{EAS}/1^{\circ})$
		}\\
		\hline
		Cluster azimuth angle & $\phi_{n_c}$ & \makecell{
		$\phi_{n_c}=X_n\phi'_{n_c}+Y_n+\phi_k$, $X_n\in\{-1,1\}$, $Y_n\sim \mathcal{N}(0,(\text{ASA}/7)^2)$ 	\\
		$\phi'_{n_c}=2(\text{ASA}/1.4)\sqrt{-\log(P_{n_c}/\max(P_{n_c}))}/C_{\phi}$, $C_{\phi}=1.273$
		}\\
		\hline
		Cluster zenith angle & $\theta_{n_c}$ & \makecell{
		$\theta_{n_c}=X_n\theta'_{n_c}+Y_n+\theta_k$, $X_n\in\{-1,1\}$, $Y_n\sim \mathcal{N}(0,(\text{EAS}/7)^2)$\\
		$\theta'_{n_c}=-\text{EAS}\log((P_{n_c}/\max(P_{n_c})))/C_{\theta}$, $C_{\theta}=1.184$	
		}\\
		\hline
		Ray power in cluster $n_c$ & $P_{n_c,n}$ & \makecell{
		$P_{n_c,n}=P_{n_c}\times P'_{n_c,n}/\sum_{n=1}^{N}P'_{n_c,n}$\\
		$P'_{n_c,n}=\exp(-\sqrt{2}\alpha_{\text{ASA}}/11)\exp(-\sqrt{2}\alpha_{\text{EAS}}/9)$,$\alpha_{\text{ASA}}, \alpha_{\text{EAS}}\sim \mathcal{U}(-2,2)$ %\max\{0.25, 6.5622-3.4084\log_{10}(f_c)\}$,
		}\\
		\hline
		Ray azimuth angle in cluster $n_c$ & $\phi_{n_c,n}$ & \makecell{
		$\phi_{n_c,n}=\phi_{n_c}+c_{\text{ASA}}\alpha_n$, $\alpha_n\sim \mathcal{U}(-2,2)$
		}\\
		\hline 
		Ray zenith angle in cluster $n_c$ & $\theta_{n_c,n}$ & \makecell{
			$\theta_{n_c,n}=\theta_{n_c}+c_{\text{EAS}}\alpha_n$, $\alpha_n\sim \mathcal{U}(-2,2)$
		}\\
		\hline
		Ray complex attenuation factor in cluster $n_c$ & $\alpha_{n_c,n}$ & \makecell{
		$\alpha_{n_c,n}=\sqrt{P_{n_c,n}}\exp(j\Phi)$, $\Phi\sim \mathcal{U}(-\pi,\pi)$
		}\\
		\hline
	\end{tabular}
\end{table*}

\subsection{Cost-efficiency and sensitivity analysis}\label{sub_sec:cost}
In contrast to the cell sectoring system based on ULA with HBF that requires $3N_\text{RF}M$ phase shifters, the cylinder DCAA only needs $N_\text{RF}N$ single port double throw (SPDT) RF switches to achieve flexible beam steering, though at the expenditure of more antenna elements arranged in the array. 
For a system satisfying scenario (ii) discussed in Section \ref{sub_sec:simu} that works under frequency $37$ GHz, we assume a mmWave communication systems that employs the TPG2102 5-bits phase shifter and the TGS4302 SPDT RF switch. 
% For the cylinder DCAA discussed in Section \ref{sub_sec:simu} of scenario (ii) working under frequency $37\enspace\text{GHz}$, we assume a mmWave communication systems that employs the ADMV4728 beamformer and the TGS4302 SPDT RF switch. 
From the manufacture's quotation, the detailed prices of the above basic hardware components are given in Table \ref{tab:cost}. 

\begin{table}[htbp]
	\centering
	\caption{Hardware components quotation for mmWave communication.}
	\label{tab:cost}
	\begin{tabular}{|c|c|c|}
		\hline
        \multicolumn{3}{|c|}{carrier frequency $f_c=37$ GHz}\\
        \hline
		\textbf{Component} & \textbf{Symbol} & \textbf{Quotation in dollar} \\
		\hline
		  \makecell{Antenna\\elements} & $c_\text{An}$ & $0.01$\\
		\hline
		  \makecell{TPG2102\\5-bits phase shifter} & $c_\text{PS}$ & $131.2$ \cite{ref:shifter}\\
		\hline
		\makecell{TGS4302\\RF switch} & $c_\text{SW}$ & $28.62$ \cite{ref:switch}\\
		\hline
	\end{tabular}
\end{table}
Accordingly, the primary hardware expenditures for the cylinder DCAA and cell sectoring based on ULA with HBF is given in Table \ref{tab:cost_sys}. 
% The total cost of the communication system with cylinder DCAA and cell sectoring system based on ULA with HBF is given in Table \ref{tab:cost_sys}. 
\begin{table}[htbp]
	\centering
	\caption{Hardware costs for two communication systems.}
	\label{tab:cost_sys}
	\begin{tabular}{|c|c|c|}
		\hline
        \multicolumn{3}{|c|}{$M=128,K=N_\text{RF}=30$}\\
        \hline
		\textbf{System} & \textbf{Expression} & \textbf{\makecell{Numerical results\\in dollar}} \\
		\hline
		  \makecell{Cylinder\\DCAA} & $c_{\text{cylin}}=2NMc_{\text{An}}+N_{\text{RF}}Nc_{\text{SW}}$ & $89,560.64$\\
		\hline
		  \makecell{ULA\\with HBF} & $c_{\text{ULA}}=3N_{\text{RF}}Mc_{\text{PS}}+3Mc_{\text{An}}$ & $1,511,427.84$\\
		\hline
		\multicolumn{3}{|c|}{$c_{\text{cylin}}/c_{\text{ULA}}=5.93\%$}\\
		\hline
	\end{tabular}
\end{table}
% For the communication system mentioned in the former section, cylinder DCAA costs about $c_{\text{cylin}}=\text{\$} {89,561}$ and cell sectoring system based on ULA with HBF costs about $c_{\text{ULA}}>\text{\$}{460,803}$. 
% Thus, we achieve better maximum sum rate performance as well as decreased hardware cost to about . 
% Table \ref{tab:cost_sys} illustrates that the cylinder DCAA achieves lower hardware costs with 
The results in Table \ref{tab:cost_sys} demostrates that the proposed cylinder DCAA can significantly reduce the hardware cost, amounting to only $5.93\%$ of the required by the cell sectoring based on ULA with HBF. 
Sensitivity analysis indicates that the cylinder DCAA performs more cost-efficiently when the cost of an individual antenna element does not exceed \$ 54.2. 
% This result highlights the economic advantage of the cylinder DCAA configuration under cost-constrained antenna design scenarios.
% Sensitivity analysis shows that the cylinder DCAA is more cost-efficient for antenna element who costs no more than $\text{\$}54.2$.

\section{Conclusion}
In this paper, to solve the signal blockage problem inherent in RAA, we propose a novel architecture termed cylinder DCAA and its optimization algorithm for uplink and downlink communication. 
First, we analyze the characteristics of sUCA, which is the basic component of cylinder DCAA. 
Second, the cylinder DCAA's geometry structure is derived and the formulated expressions of its response pattern, uplink and downlink SINR and maximum sum rate are presented. 
Compared with ULA with HBF, cylinder DCAA has uniform spatial resolution, enhanced beamforming gain with directional antenna elements and substantial cost-efficiency. 
Based on the characteristics of cylinder DCAA, optimization algorithms to find the maximum sum rate in the uplink and downlink with greedy method or iteration-based method are proposed. 
% and the numerical results have demonstrated the effectiveness and convergence of the proposed algorithms. 
%The theoretical analysis for the volume of stacked polyprism architecture which is one of the potential RAA implementation and the cylinder architecture shows that cylinder architecture always has smaller space occupation with the same number of antenna elements in a subarray. 
%Thus, cylinder architecture is more potential for the UAV communication scenario. 
%The near-field and far-field boundary of cylinder architecture is a circle that is independent of AoA set $(\phi,\theta)$ in direction-dependent Rayleigh distance for reference. 
Finally, the numerical results validate that cylinder DCAA has better sum rate performance than cell sectoring system based on ULA with HBF, especially in dense connectivity scenario, and the proposed algorithms are effective in finding the optimal power allocation with practical computational complexity together with good convergence performance.  
These inspirational findings demonstrate that the cylinder DCAA is a potential solution for the signal blockage problem of RAA. 
% typical dense connectivity in 6G mmWave or THz communcation. 
In the further work, we can explore lower complexity and more efficient algorithms for cylinder DCAA and some strategies of integrated sensing and communication (ISAC) for the communication system with cylinder DCAA. 

{\appendices
	
\section{proof of theorem \ref{the:AF}}
\label{app:proof2}

The amplitude of array factor $\text{AF}(\eta,\phi,\theta)$ in \eqref{AF_temp} is equivalently expressed as summarized weighted uniform complex vectors given by:
\begin{equation}
    \vert\text{AF}(\eta,\phi,\theta)\vert=\left\vert\sum_{m=1}^{M}\chi_m(\phi,\eta)\mathbf{u}_m(\eta,\phi,\theta)\right\vert,
    \label{equ:app_af_sim}
\end{equation}
where $\chi_m(\phi,\eta)=\sqrt{G(\xi_m(\eta),\psi)}$, $\mathbf{u}_m(\eta,\phi,\theta)=e^{jx_m(\eta,\phi,\theta)},\quad (m=1,2,\cdots,M)$ and $x_m(\eta,\phi,\theta)$ is given by:
\begin{equation}
\begin{aligned}
        &x_m(\eta,\phi,\theta)\\
        &=\frac{4\pi}{\lambda}a\sin\theta\cos\underbrace{\left(\rho(\phi,\eta)-\frac{\pi}{M-1}(m-1)\right)}_{\zeta(m,\phi,\eta)}\sin(\rho(\phi,\eta)),
\end{aligned}
    \label{equ:app_x}
\end{equation}
where $\rho(\phi,\eta)=\frac{\phi-\eta}{2}$. 
Since $\rho(\phi,\eta)-\pi\leq\zeta(m,\phi,\eta)\leq\rho(\phi,\eta)$, the supremum and the infimum of the phase set relative to the index $m$ is given by:
\begin{equation}
    \begin{aligned}
        &\underset{1\leq m\leq M}{\mathbf{sup}}\{x_m\}
        \approx\left\vert\frac{4\pi}{\lambda}a\sin\theta\sin\left(\rho(\phi,\eta)\right)\right\vert>0\\[8pt]
        &\underset{1\leq m\leq M}{\mathbf{inf}}\{x_m\}
        =-\left\vert\frac{4\pi}{\lambda}a\sin\theta\cos\left(\rho(\phi,\eta)\right)\sin\left(\rho(\phi,\eta)\right)\right\vert<0.\\
    \end{aligned}
    \label{equ:app_phase}
\end{equation}
From \eqref{equ:app_phase}, the infimum of the set is negative and the supremum of the set is positive. 
Thus, there always exists a specific $1\leq i_k\leq M$, such that $\zeta(i_k,\phi,\eta)\approx-\frac{\pi}{2}$. 
Based on the symmetric property of $\cos(\cdot)$ at the neighborhood of zero point, for a given angle set $(\phi,\theta)$, $\phi\in\left[\eta-\frac{\pi}{2},\eta+\frac{\pi}{2}\right]$ and $\theta\in[0,\pi]$, there exists a sub-set of $2k$ vectors $\{\mathbf{u}_{i_1},\mathbf{u}_{i_2},\cdots,\mathbf{u}_{i_{2k}}\}\subseteq\{\mathbf{u}_m\}$ that are symmetric relative to the real axis, i.e., $\angle\mathbf{u}_{i_j}+\angle\mathbf{u}_{i_{2k-j+1}}=0, j=1,2,\cdots,2k$, where $M/2\leq2k\leq M$. 
Thus, complex uniform vector set is partitioned into two parts: 
\begin{equation}
    \begin{aligned}
        &\{\mathbf{u}_m\enspace\vert\enspace m=1,2,\cdots,M\}\\
        &=\{\underbrace{\mathbf{u}_{i_1},\mathbf{u}_{i_2},\cdots,\mathbf{u}_{i_{2k}}}_{\text{symmetric part}},\underbrace{\mathbf{u}_{i_{2k+1}},\mathbf{u}_{i_{2k+2}},\cdots,\mathbf{u}_{i_{M}}}_{\text{non-symmetric part}}\}.
    \end{aligned}
    \label{equ:app_sub_set}
\end{equation}
Thus, \eqref{equ:app_af_sim} is equivalently expressed to:
\begin{equation}
    \begin{aligned}
        &\vert\text{AF}(\eta,\phi,\theta)\vert\\
        &=
        \vert\underbrace{\sum_{j=1}^k(\chi_{i_j}\mathbf{u}_{i_j}+\chi_{i_{2k-j+1}}\mathbf{u}_{i_{2k-j+1}})}_{\triangleq\alpha}+\underbrace{\sum_{j=2k+1}^{M}\chi_{i_j}\mathbf{u}_{i_j}}_{\triangleq\beta}\vert\\
        &=\vert\alpha-\beta^*+\beta+\beta^*\vert\overset{(a)}{\leq}\vert\alpha-\beta^*\vert+\vert\beta+\beta^*\vert, 
    \end{aligned}
    \label{equ:app_ori}
\end{equation}
where $(a)$ is based on Jensen's inequality.
The maximum of \eqref{equ:app_ori} can only be obtained when $\angle(\alpha-\beta^*)=0$. 
However, with the existence of non-symmetric vectors, i.e., $\phi\neq\eta$ and therefore $\beta^*\neq0$, it is intuitive that $\angle(\alpha-\beta^*)\neq0$. 
Thus, the maximum of \eqref{equ:app_af_sim} is obtained only at angle set $(\phi,\theta)=(\eta,\frac{\pi}{2})$. 
From \eqref{equ:valleypoint}, the beamwidth $\text{BW}=2\phi_0\approx2\tilde{\phi}_0$. 
Thus, the proof of \emph{Theorem} \ref{the:AF} is completed.

}

\bibliography{IEEEabrv,myrefs}

\end{document}